\documentclass[%
 reprint,
 amsmath,amssymb,
 aps,
]{revtex4-2}

\usepackage{graphicx}%
\usepackage{dcolumn}%
\usepackage{bm}%
\usepackage{wasysym}

\usepackage{titlesec}
\titlespacing{\paragraph}{%
  0pt}{%
  0.5\baselineskip}{%
  1em}
\titleformat{\paragraph}[runin]
{\bfseries\itshape}{\theparagraph}{1em}{}

\usepackage{LN_Macros}

\begin{document}

\preprint{APS/123-QED}

\title{Empirical Sampling of Connected Graph Partitions for Redistricting}%

\author{Lorenzo Najt}
 \email{lnajt@math.wisc.edu}
\affiliation{%
Department of Mathematics, University of Wisconsin.}%

\author{Daryl DeFord}
 \email{daryl.deford@wsu.edu}
 \affiliation{Department of Mathematics and Statistics, Washington State University}
 
 \author{Justin Solomon}
  \email{jsolomon@mit.edu}
\affiliation{
 Computer Science and Artificial Intelligence Laboratory, Massachusetts Institute of Technology
}%

\date{\today}%

\begin{abstract}
The space of connected graph partitions underlies statistical models used as evidence in court cases and reform efforts that analyze political districting plans. In response to the demands of redistricting applications, researchers have developed sampling methods that traverse this space, building on techniques developed for statistical physics.  
 In this paper, we study connections between redistricting and statistical physics, and in particular with self-avoiding walks. We exploit knowledge of phase transitions and asymptotic behavior in self avoiding walks to
 analyze two questions of crucial importance for Markov Chain Monte Carlo analysis of districting plans. First, we examine mixing times of a popular Glauber dynamics based Markov chain and show how the self-avoiding walk phase transitions interact with mixing time. We examine factors new to the redistricting context that complicate the picture, notably the population balance requirements, connectivity requirements, and the irregular graphs used. Second, we analyze the robustness of the qualitative properties of typical districting plans with respect to score functions and a certain lattice-like graph, called the state-dual graph, that is used as a discretization of geographic regions in most districting analysis. This helps us better understand the complex relationship between typical properties of districting plans and the score functions designed by political districting analysts. %
 We conclude with directions for research at the interface of statistical physics, Markov chains, and political districting.
\end{abstract}

\maketitle

\section{Introduction}
\label{sec:intro}

Several recent court cases and reform efforts have created demand for quantitative techniques to analyze \emph{political districting plans}, which are divisions of a state or municipality into regions that each elect a candidate to office.  Reflecting political practice, districting plans are considered as connected partitions of geographic subunits, e.g., voting precincts or census blocks.  Since the space of all possible such partitions is too large for complete enumeration, most analyses rely on random sampling to generate an ensemble of plans for a given piece of geography; the properties of this ensemble are used to understand the achievable qualities of potential plans and inevitable trade-offs between the many objectives for plan design.

An increasingly popular methodology for sampling districting plans employs  Markov Chain Monte Carlo (MCMC) techniques. %
A representative subset of examples includes \cite{calderamathematics,chikina_assessing_2017,deford_compete,deford2019redistricting,Fifield_A_2018,herschlag_quantifying_2018,herschlag_evaluating_2017,deford2019recombination}; see \cite{algs_chapter} for broader discussion of methods in computational redistricting.  Although districting plan samplers have been applied convincingly in several court cases \cite{chen1,herschlag_quantifying_2018,herschlag_evaluating_2017, pegden1}, these arguments relied on sampling methods manually tuned to the geographies and legal constraints of specific states, suggesting the need for further scientific research into this potentially useful tool.

To enhance understanding of these MCMC methods, in this paper we draw connections between political districting and statistical physics. We will see that statistical physics helps discover and explain the behavior of sampling methods used in redistricting.  In the reverse direction, redistricting motivates new variants of commonly-studied statistical physics models, including new questions that are especially natural in the districting context. %

One of the popular Markov chains for districting explores the space of districting plans through a variant of Glauber dynamics. However, a distinguishing feature between redistricting and traditional spin glass models is that districting plans are required to be \emph{contiguous}. That is, the nodes assigned to each label must induce connected subgraphs. This constraint connects the behavior of redistricting methods to known results about self-avoiding walks, but makes it more difficult to use intuitions from spin glass models. 
In particular, fast mixing properties of cluster-based spin glass models over portions of the parameter space \cite{huber02,huber99,park17,perez11} do not appear to hold generically for the samplers we consider. For instance, \cite{najt2019complexity} demonstrates general obstructions to efficient uniform sampling from the space of connected districting plans, along with explicit families of graphs where the Glauber dynamics based Markov chains on districting plans mix slowly. However, the behavior of these Markov chains on regular lattices has so far resisted analysis; in this paper, we provide some empirical evidence regarding the behavior of these chains on lattices.

Moving beyond contiguity, other factors such as population balance and irregularity of the underlying graph present  significant departures from physical models. These complications affect mixing times and qualitative properties of stationary distributions; however, many facts about phase transitions and lattice dependencies carry over to the redistricting context. %

We also encounter some questions not explicitly studied in statistical physics, since they have to do with the specific way that samplers are used in redistricting. In the redistricting analysis pipeline, a finite graph representing the state is constructed, called the state-dual graph (see \Cref{sec:the_redistricting_problem}), and the aforementioned MCMC samplers are used to produce partitions of that graph that correspond to legal districting plans. We demonstrate, using knowledge of phase transitions gleaned from statistical physics, that the way that the graph is constructed can influence the analysis, even as the parameters of a sampling algorithm are held constant. %
Such examples raise important questions about the practical interpretation of statistics about random redistricting plans.

A theory for resolving the problem of \emph{which} distributions over partitions are appropriate for redistricting analysis has thus far escaped analysis, or even a precise formulation of the basic questions. We hope that progress on this question can be made by strengthening connections with statistical physics.

\section{Background on redistricting}\label{sec:the_redistricting_problem}

    \begin{figure}
        \centering
        \begin{tabular}{c}
        \includegraphics[scale = .15]{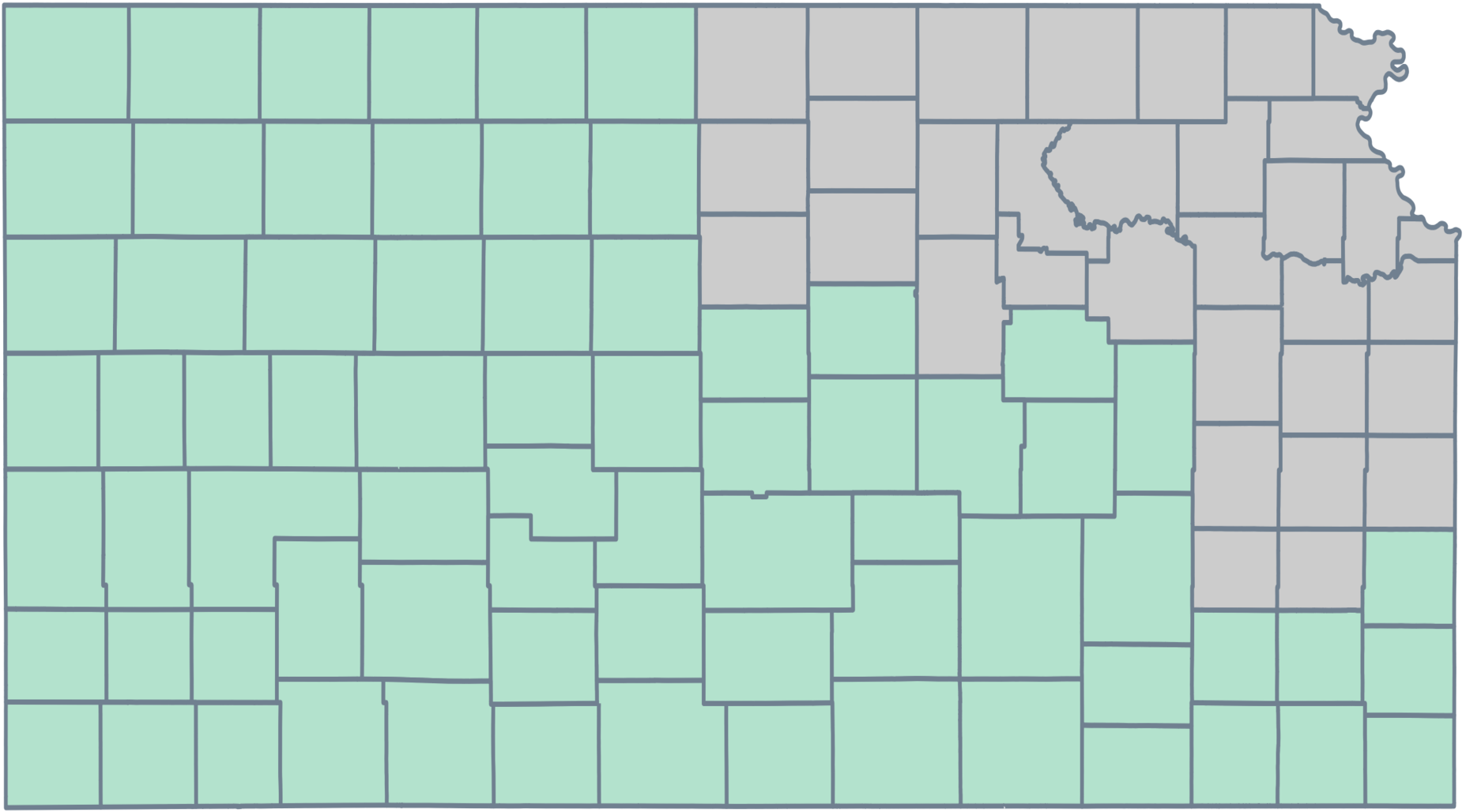} \\
        a)\\
        \includegraphics[scale = .15]{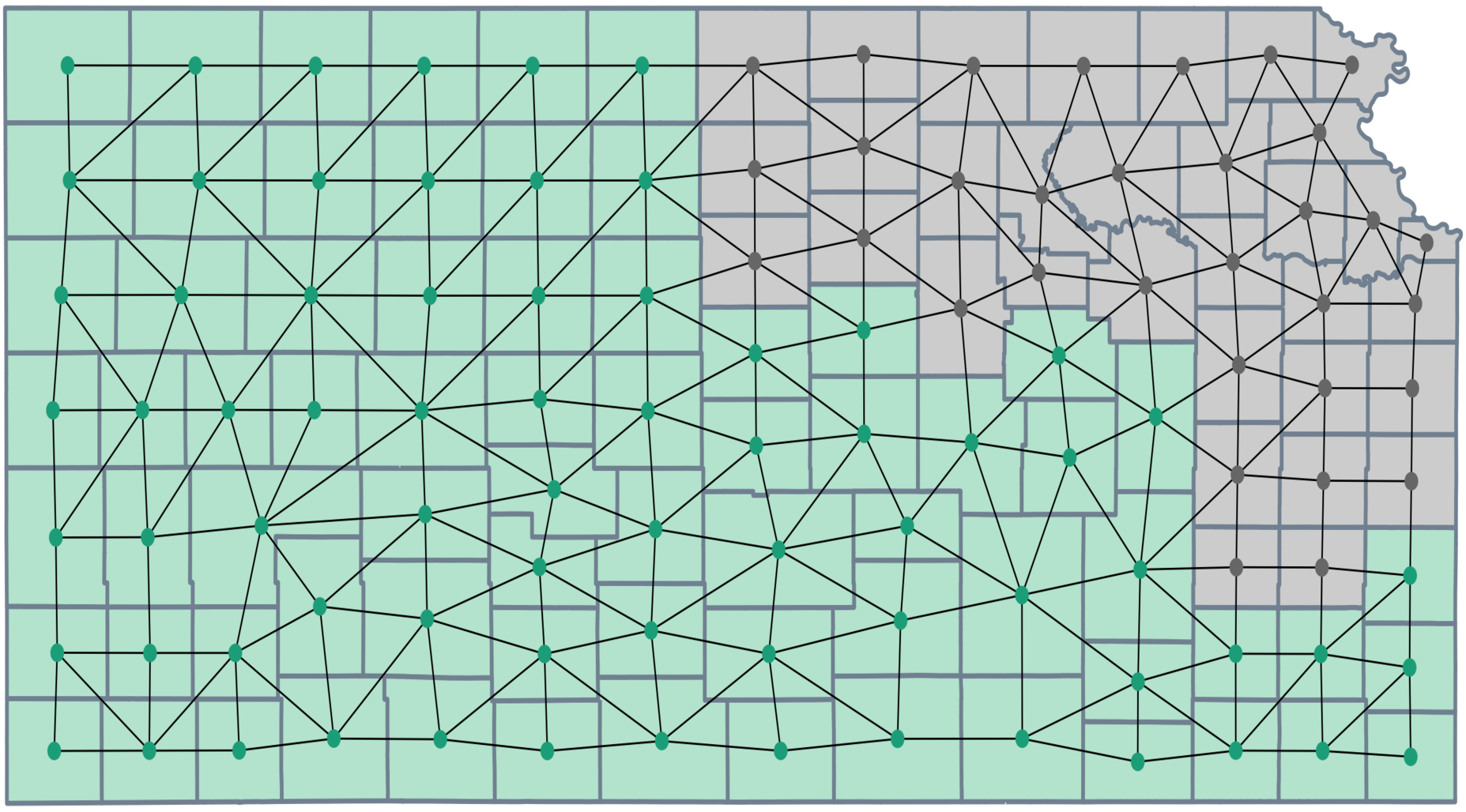}\\b)
        \end{tabular}
        \includegraphics[height=.5in]{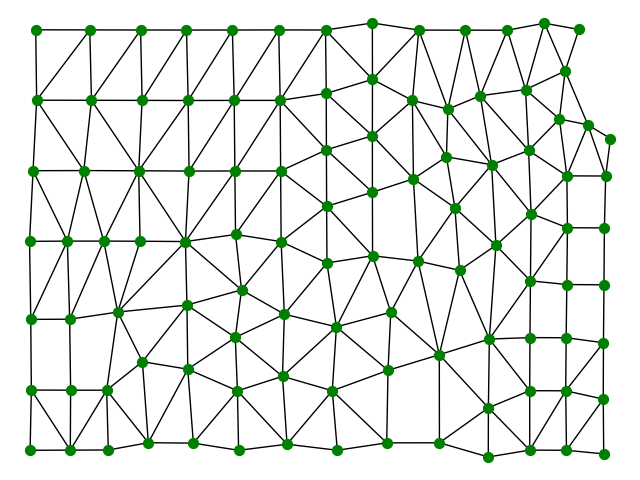}
        \includegraphics[height=.5in]{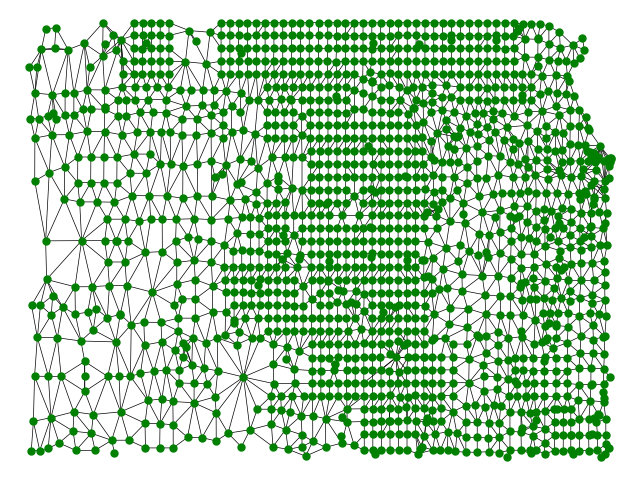}
        \includegraphics[height=.5in]{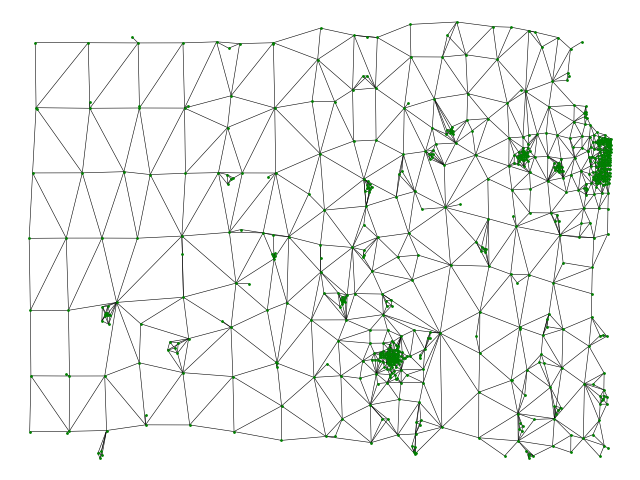}
        \includegraphics[height=.5in]{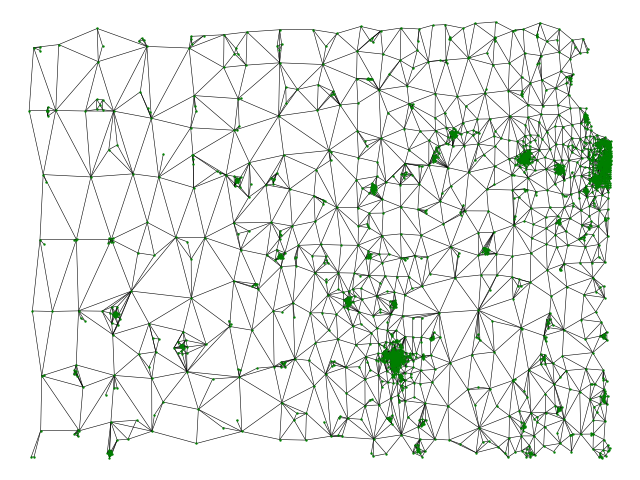}
        \caption{a) Kansas with county units \cite{shapefiles}, along with a connected $2$-partition. b) The corresponding state dual graph overlayed. Much more granular subdivisions of a state such as the census units shown in the bottom row are often used instead of counties. %
        }\label{fig:kansas}
    \end{figure}

In a typical redistricting process, states are partitioned into small units such as census blocks or voting precincts for data recording, which are then aggregated into larger districts in which political representatives are elected. The units can be represented by the nodes of a graph, where units that share common boundaries are adjacent, as in \Cref{fig:kansas}. This graph is called the \emph{state dual graph}. Assuming that voting districts must be contiguous, a districting plan with $k$ districts is modeled by a connected $k$-partition of the state dual graph. Each state has its own rules for determining which partitions are permissible, although these statutes rarely provide sufficient detail to provide a formal mathematical description of the space of politically valid plans. For more details on the problem setup, see \cite[\S3]{deford2019recombination}.

For our purposes, two of these constraints are particularly relevant. The first is population balance, which requires that the population within each of the $k$ districts be near the the state's total population divided by $k$. The second is the notion of geographic compactness, which refers to the `niceness' of district boundaries. Compactness is  evaluated using a variety of geographic  metrics that are designed to compare the shape of a given district to some idealized geometric property.  One of the most common is called the Polsby--Popper score \cite{pp}, which measures the isoperimetric ratio of perimeter to area of each district. Other measures compute ratios of areas to regularized bounding shapes or measure the length of the perimeter directly. In this paper we use a family of scores that measure the compactness of a districting plan as a function of the number of edges in the state dual graph going between different districts. This score is compelling because of its mathematical convenience and relationship to the statistical physics literature, and it has also received some attention in the redistricting literature \cite{cohen2020computational, deford2019recombination}.

It was quickly observed \footnote{See \cite{martis2008original, hunter2011first} for historical accounts.} that by a clever choice of districting plan one could engineer aspects of electoral outcomes, a practice that became known as ``gerrymandering.'' In an effort to counteract this behavior, there have been many proposals to design districting plans algorithmically, a process which often involves grappling with computationally intractable problems \cite{kueng2019fair,  altman1997automation}. The reality of redistricting, however, is that the power to draw the graph partition is in the hands of a legislature, dedicated committee, or hired expert---rather than a piece of software. For this reason, rather than using an algorithm to draw plans in the first place, some have suggested to analyze already drawn plans for compliance with civil rights law and other redistricting principles.

One approach to this analysis tries to understand a plan in the context of what alternative plans are possible or typical. For instance, an assertion that it is impossible to provide a certain number of majority-minority districts can sometimes be falsified by a computer search for such plans, as in \cite{chen_brief}. However, since the space of districting plans is in general too large to be completely enumerated it is generally computationally intractable to prove that certain kinds of plans are impossible. There are exceptions, however: one can refer to \cite{fifield2020essential} for enumerations of medium size state dual graphs and \cite{chen_brief} for situations where impossibility arguments based on non-geographic constraints are possible. Additionally, in many cases the relevant questions have to do with how unusual a plan is along a certain dimension, rather than the impossibility or possibility of a certain kind of plan.

As such, appeals to notions of a \emph{typical} plan and statistical outliers have developed into popular lines of argument. For instance, an argument that a plan was drawn with the intent to discriminate might calculate that the proposed plan has more discriminatory properties than the vast majority of plans from a randomly generated collection of comparable plans, as was done in an amicus brief that was presented to the Supreme Court of the United States in 
Rucho v.\ Common Cause \cite{mathematiciansbrief}. The arguments given in that amicus brief relied on a choice of distribution over the space of plans and attempted to tailor that distribution to sample from a \emph{diverse} \emph{ensemble} of plans that are each compliant with the redistricting principles laid out by the governing body. %
More generally, a variety of algorithms have been proposed to sample ensembles of graph partitions for similar purposes, from genetic algorithms \cite{cho_toward_2016, liu_pear:_2016} to random walks \cite{chikina_assessing_2017, herschlag_quantifying_2018}; recent expert reports in redistricting cases have used these tools to generate quantitative assessments of proposed plans \cite{chen1,herschlag_quantifying_2018,herschlag_evaluating_2017, pegden1}. 

In this paper, we focus on algorithms based on Markov chain Monte Carlo, due to a connection with the statistical physics of random spanning trees, self-avoiding walks, and phase transitions in Gibbs energies. Methods making use of these tools have already found wide application in court cases and academic analyses \cite{mathematiciansbrief, herschlag_quantifying_2018,herschlag_evaluating_2017, pegden1}. We will concern ourselves primarily with questions about the mixing time of these Markov chains, and the dependence of the properties of typical plans on phase transitions in the stationary distributions. For instance, we will find that the phase transitions are useful for understanding how a redistricting analysis depends on the discretization of the geographic region into a state dual graph, which is an important point for developing ``objective'' (in the sense of \cite{gelman2017beyond}) standards for restricting analysis. Phase transitions in a Potts model related to these redistricting problems are considered from a statistical physics perspective in \cite{zohar}.

\subsection{Two key questions for MCMC samplers in redistricting}

Two critical challenges arise when using statistics of random districting plans to analyze political redistricting. First, when the random plans are produced by a Markov chain, we would like to estimate the computation time required to produce a statistically representative sample from the stationary distribution of that chain. Second, to make the analysis more \emph{transparent} and \emph{impartial} (in the sense of \cite{gelman2017beyond}), we would like to characterize how the statistics used in the analysis are influenced by features of the sampling algorithm.

This second challenge is especially subtle in the context of the data analysis pipeline for political redistricting, since an outlier classification of a districting plan could be affected by ad-hoc seeming features of the discretization of the state into a state dual graph, as we explore in \cref{sec:role_of_discretization}. %
Additionally, although explicit score functions can be designed for use with the Metropolis-Hastings algorithm, such as those in \cite{carter2019merge,mattingly1, herschlag_quantifying_2018,Fifield_A_2018}, 
there is often a significant difference between the apparent preferences of a score function and the typical properties of a sample from the corresponding stationary distribution. %
Concepts from statistical physics, such as phase transitions around lattice-dependent constants, are key for understanding the relationship between designed score functions and the qualitative properties of typical samples.

In this paper, we will use those physical concepts to build evidence for conjectures about mixing times and to investigate the robustness of redistricting analysis with respect to the state dual graph. We believe that our investigation provides strong evidence that insights from statistical physics are valuable for developing best practices in redistricting analysis. %

\section{Metropolis--Hastings based samplers}

We recall that a general paradigm for Metropolis--Hastings based sampling requires that we build a reversible Markov chain $M$ over the state space $\Omega$  of interest and define an energy function $E : \Omega \to \mathbb{R}$ proportional to the desired stationary distribution. The Metropolis--Hastings algorithm uses the energy function to modify the proposal of $M$ in the following way: if the proposal chain suggests moving to a state of lower energy, the Metropolis--Hastings chain always accepts. On the other hand, if the proposal chain suggests moving to a state of higher energy, the Metropolis--Hastings chain only accepts with probability $\exp ( E(x_{\text{current}}) - E(x_{\text{proposal}}) )$. 

In redistricting, a popular approach has been to define an energy function that incorporates many of the relevant criteria, such as geographic compactness (see \Cref{sec:the_redistricting_problem}), population balance, county preservation, and VRA compliance, among others  \cite{carter2019merge,mattingly1, herschlag_quantifying_2018,Fifield_A_2018,mcmc_chapter, deford2019recombination}. Analysis depending on such Markov chains have appeared in expert reports in federal court cases, as such as those discussed in \cite{mattingly1, pegden1, mathematiciansbrief}. We explain these choices in further detail in \Cref{sec:gibbs}. As is typical of many such systems, there is a tug of war between combinatorial explosion and the energy minimization resulting in phase transitions in the tuning of these energy functions; these phase transitions will be a central part of the story we tell in \Cref{sec:phasetransitions}.

In general, one can use the Metropolis--Hastings scheme over a larger state space, say $\Omega'$, than the one, $\Omega$, that one is explicitly interested in, and then rejection sample to obtain samples in $\Omega$. This approach is useful if enlarging the state space makes the Markov chain mix more quickly and the probability that the stationary distribution assigns to $\Omega$ is not negligible. This approach can be useful for certain features of redistricting plans, such as population balance. However, in general those interested in analyzing redistricting plans have found it convenient to preserve the connectedness requirement in the proposal function; thus, we will be focusing on proposals that return a connected partition at each step of the chain. 

There are two predominant proposal types used in redistricting, one is a Glauber dynamics like proposal based on node flips and the other is a family of proposals based on using spanning trees as a certificate of connectedness. We discuss these in more detail in the next section.

\subsection{Two proposal methods for connected partitions}\label{sec:proposals}

In this section, we define connected partitions and describe two types of Markov chains on connected partitions.

\begin{defn}[Connected Partitions]
Let $G$ be a graph. A connected partition of $G$ is a partition of the vertices such that each block induces a connected subgraph of $G$. We let $P_k(G)$ denote the set of connected, ordered $k$-partitions; equivalently, $P_k(G)$ is the set of $k$ colorings of $G$ where each coloring induces a connected subgraph. We call such a coloring a connected $k$-coloring.
\end{defn}

The first Markov chain we define is called the flip walk, which is a variant on Glauber dynamics:

\begin{defn}[Flip Walk]
Let $G = (V,E)$ be a graph. We define a Markov chain on the state space $P_k(G)$ by defining a step from a connected coloring $P$ by picking a $v \in V$ uniformly at random, and a $c \in [k]$ uniformly at random, and proposing a move to the coloring $P'$ with $P'(w) = P(w)$ for $w \not = v$, and $P'(v) = c$. If $P'$ is a connected coloring, this move is accepted, otherwise it is rejected.
\end{defn}

Additional discussion of flip variants can be found in \cite[\S4.2]{deford2019recombination}. 

\begin{remark}
In \Cref{sec:phasetransitions} we describe the connection between 2-connected partitions of the grid and self-avoiding walks. The flip walk on connected 2-partitions can thus be translated over to self-avoiding walks, and results in a chain that is essentially identical to the BFACF walk \cite{sokal1994monte}. See \cite{najt2019complexity} for a literature review.
\end{remark}

Another MCMC algorithm for redistricting is based on spanning trees:

\begin{defn}[Recombination]
Let $G = (V,E)$ be a graph. We define a Markov chain on the state space $P_k(G)$ by defining a step from a connected coloring $P$ by picking two adjacent colors at random and drawing a random spanning tree $T$ for the subgraph of $G$ induced on the union of the nodes of those two colors. We then randomly select an edge of $T$ to cut, leaving behind two connected components, which are randomly assigned one of the two originally selected colors.  
\end{defn}

The recombination algorithm was introduced to the redistricting literature in \cite{deford2019recombination}, which contains a discussion of several variants of the method depending on the algorithms used in the randomized parts of the procedure, such as selecting the initial colors, the spanning tree, or the edge to cut. However, the idea of using spanning trees as a certificate for connectedness is much older and pervasive throughout combinatorics and statistics \cite{lovasz1977homology, meila2018sample, vo2019dimension, validi2020imposing}. Recently, this approach has been generalized \cite{carter2019merge} to operate on the extended state space of spanning forests.

For the most part, our empirical investigation in this paper will focus on the flip walk, but we will return to the $2$-partition version of recombination in \Cref{sec:role_of_discretization}.

\subsection{Gibbs energy}\label{sec:gibbs}

Given one of these proposal mechanisms, the Metropolis--Hastings algorithm can be used to construct a Markov chain whose stationary distribution is proportional to a Gibbs energy:
\begin{equation}
    \mathbb{P}(P) = \dfrac{e^{-\beta(\sum_{i} \alpha_i \sigma_i(P))}}{Z}
\end{equation}
where $P$ is a partition of the given graph, $\beta$ is the inverse temperature,  the $\sigma_i$ are functions measuring legislatively relevant quantities of $P$ with corresponding weights $\alpha_i$, and $Z$ is the corresponding partition function:
\begin{equation}
    Z= \sum_{P\in \mathcal{G}} e^{-\beta(\sum_{i} \alpha_i \sigma_i(P))}.
\end{equation} 

Common measurements represented by the $\sigma_i$ include population balance, geographic compactness (see \Cref{sec:the_redistricting_problem}), preservation of county boundaries, and compliance with the Voting Rights Act of 1965 among others. As the relevant legislation is rarely written in sufficient detail to specify a single $\sigma$ much less an entire distribution, there is a significant amount of modeling freedom in the construction of such an energy. %

As an example, \cite{herschlag_quantifying_2018} uses simulated annealing with the following energy function to sample Congressional districting plans in North Carolina: 
\begin{equation}
\scalebox{1}{$\displaystyle{e^{-\beta \!\cdot\! \left(3000\ \sigma_{\rm pop}+2.5\ \sigma_{\rm compact}+0.4\ \sigma_{\rm county}+800\ \sigma_{\rm VRA}\right)},}$}
\end{equation}
where, in addition to terms measuring the population balance and compactness, the energy function incorporates an entropy-based measure of county splits and a term incentivizing districts with Black Voting Age Population percentages at least as large as those permitted by previous legal challenges \cite{mcmc_chapter}.

\section{Phase transitions}\label{sec:phasetransitions}

In this section, we consider a  specific score function for Metropolis--Hastings on the flip walk; namely, for a fixed $\lambda > 0$, we define $\beta_{\lambda}(P) = \lambda^{\cut(P)}$, where $\cut(P)$ is the number of edges connecting the different connected components of $P$. We explain a connection to theory of self-avoiding walks and how the phase transition behavior for self-avoiding walks carries over to the case of connected partitions and the flip walk. Following the self-avoiding walk literature, we will refer to the term $\lambda$ in $\beta_{\lambda}(P)$ as the \emph{fugacity}. Later on, we will use our knowledge about phase transitions and other self-avoiding walk theory to guide an empirical investigation of bottlenecks in the flip walk chain. To begin with, we introduce a toy model of a state dual graph that will serve as a running example throughout.

\begin{figure}
    \centering
    \begin{tabular}{c}
    \includegraphics[scale = .15]{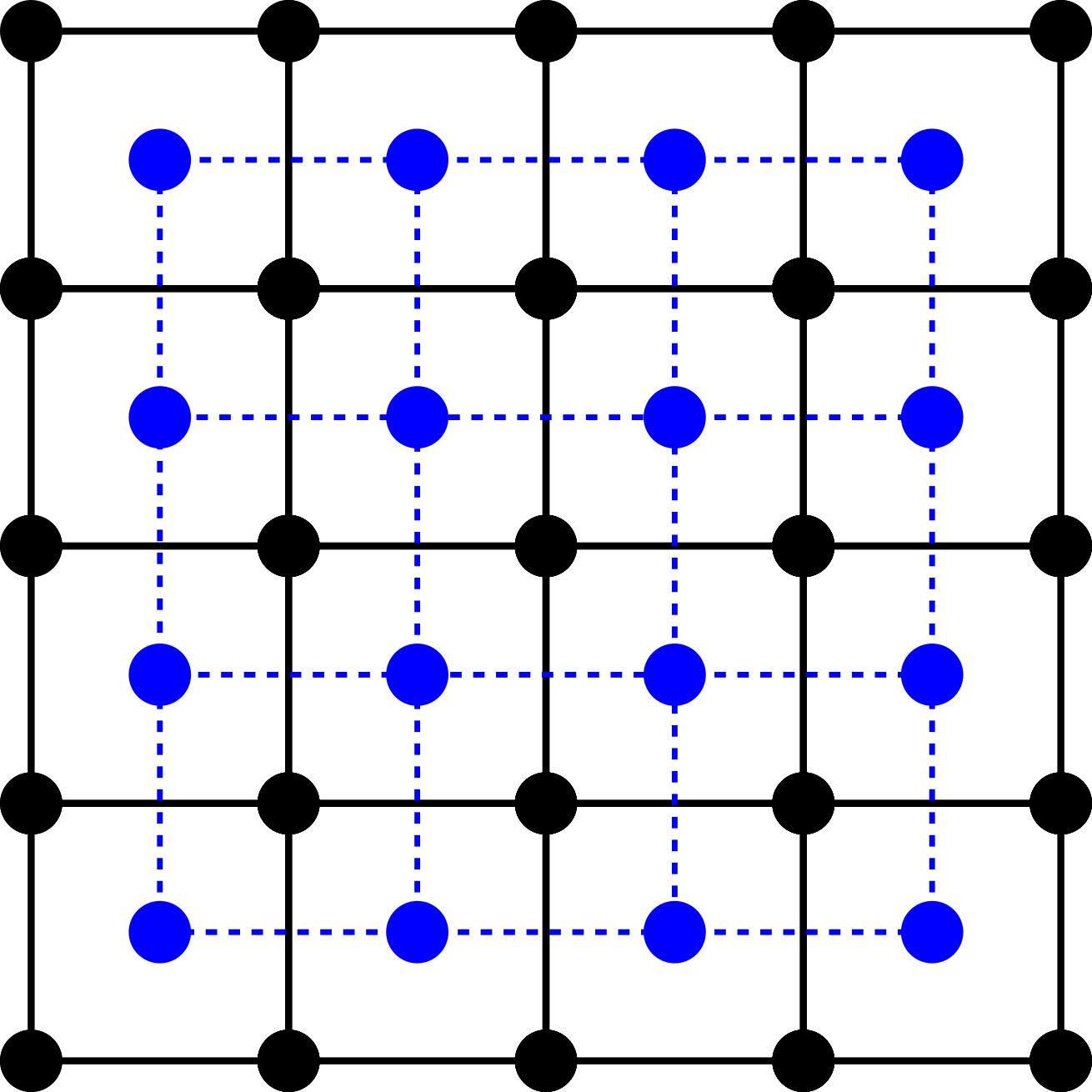} \\
        \reflectbox{\includegraphics[scale = .15, angle = 90]{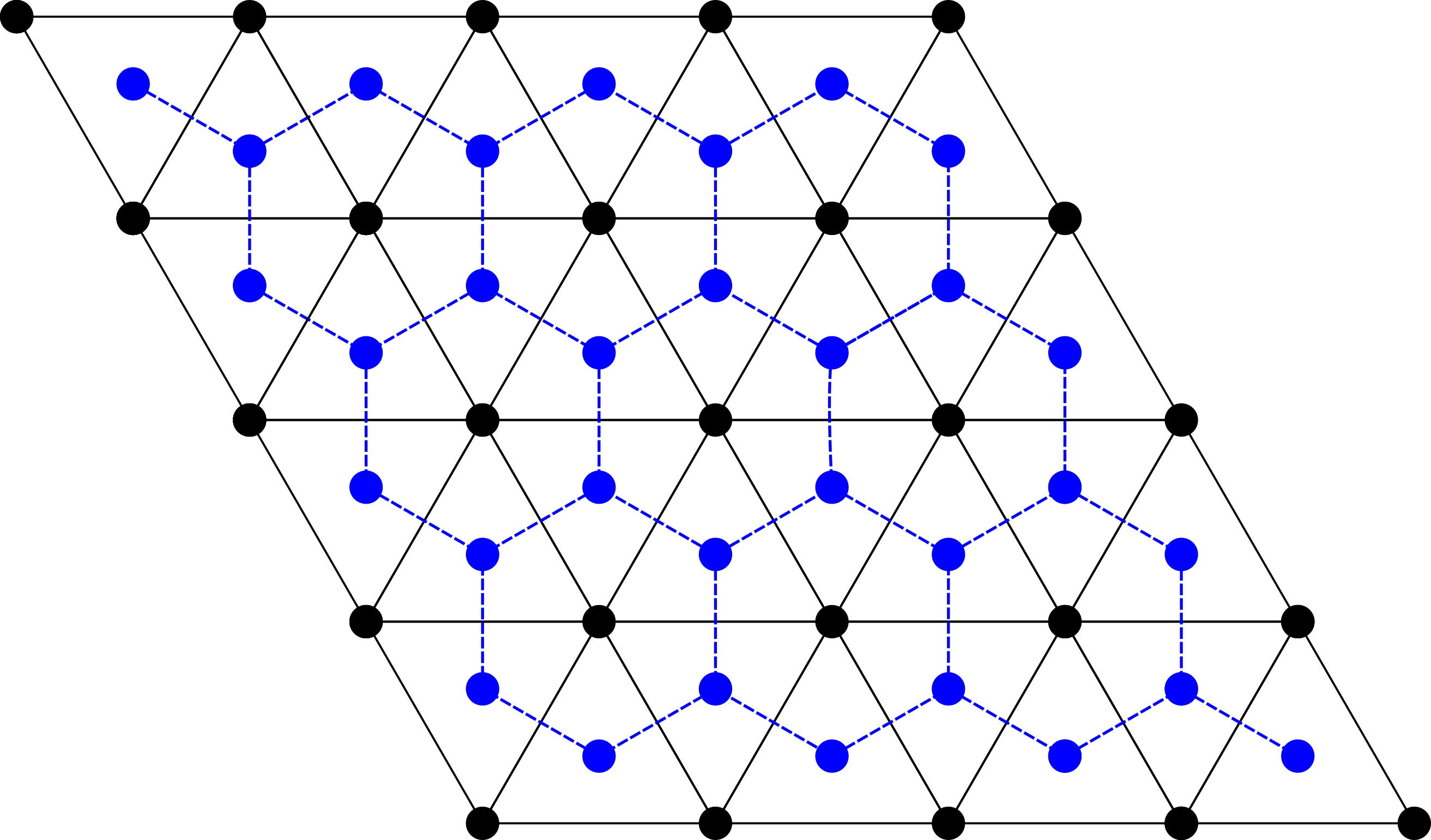}} \\\includegraphics[scale = .15]{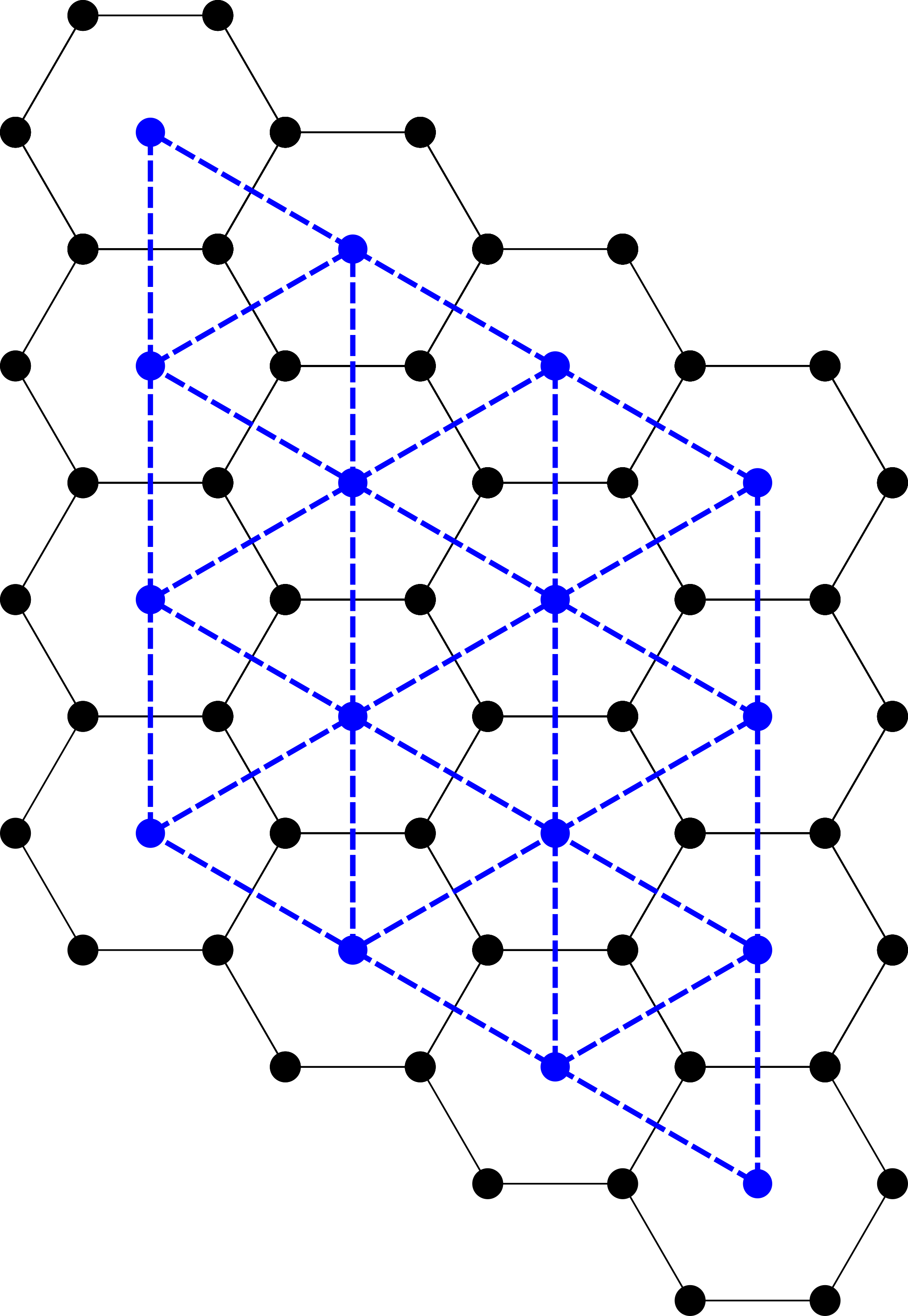}
    \end{tabular}

    \caption{From top to bottom, the $4 \times 4$ versions of Squarelandia, Trilandia and Hexlandia in black, displayed with their supernode-deleted duals in blue. }
    \label{fig:landias}
\end{figure}

\begin{defn}[Gridlandia]\label[defn]{def:gridlandia}
Gridlandia, $G_n$, is an $(n + 1) \times (n+1)$ grid graph where each node has population $1$. In other words, it is a arrangement of $n \times n$ squares as in \Cref{fig:landias}. We think of population as a function of $\mathbb{N}$-valued weights on the vertices of a state dual graph. 
\end{defn}

\begin{defn}[Population Balance]
We consider $\alpha$-balanced connected partitions of $G_n$, $P_{2, \alpha}(G_n)$, which are the connected $2$-partitions of $G_n$ where the total population of each district is within $100\alpha$ percent of exactly half the total population.%
\end{defn}

Note that increasing $\alpha$ amounts to loosening the population constraint; if $\alpha \leq \alpha'$, then $P_{2, \alpha} \subseteq P_{2, \alpha'}$.

\begin{remark}
The definition of population balance used in \Cref{def:gridlandia} is just one of many reasonable definitions. The choice of function measuring the population balance does not appear to have a significant effect on the empirical behavior. %
It is possible that another definition would be analytically preferable.
\end{remark}

\begin{figure}
    \centering
    \includegraphics[scale = 1.3]{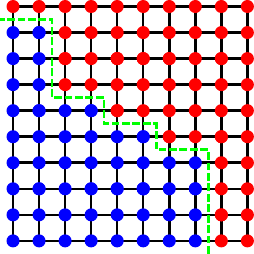}
    \caption{The relationship between connected partitions and self-avoiding walks.}
    \label{fig:SAWGridlandia}
\end{figure}

We will refer to a a non-self intersecting path between any two boundary points of $G_n$ as a \emph{chordal self-avoiding walk}. %
Using bond-cycle duality for plane graphs, we can interpret connected partitions of $G_n$ as simple cycles on the dual graph. For connected and simply connected partitions, these are effectively the same thing as chordal self-avoiding walks in the dual graph minus the node corresponding to the unbounded face, see \Cref{fig:SAWGridlandia}; the only minor discrepency being that each of the four corner faces has two edges connecting it to the supernode. Given a connected and simply connected 2-partition $P$, we refer to the boundary self-avoiding walk as the \emph{partition boundary} of $P$. 

An important family of distributions on chordal self-avoiding walks in a plane graph is the following: For any $\lambda > 0$, define the probability measure $\beta_{\lambda}(\omega) \propto \lambda^{| \omega|}$, where $|\omega|$ is the number of edges used in the self-avoiding walk. On the grid graph, the limiting behavior of these distributions experience a phase transition at $1 / \mu_S$ \cite{duminil2014supercritical}, where $\mu_S = 2.635\ldots$ is the \emph{connective constant} of the grid lattice, which measures the rate of growth of the number of length $n$ self-avoiding walks starting at the origin in $\mathbb{Z}^2$. This phase transition leads to three regimes of behavior: $\lambda < 1/\mu_S$ (subcritical), $\lambda = \mu_S$ (critical) and $\lambda > 1/\mu_S$ (supercritical). Between each regime, the behavior of a typical self-avoiding walk is very different \cite{duminil2014supercritical}, although within regimes the behavior has a qualitative uniformity. See  \Cref{sec:metastableslopes} for a discussion the properties of typical partitions sampled in these three regimes.

Connective constants for triangular and hexagonal lattices, denoted $\mu_T$ and $\mu_H$, are defined similarly to $\mu_S$, and stationary distributions for self-avoiding walks exhibit identical phase transitions for $\lambda$ at $1/\mu_L$, where $L$ is the lattice. It is known that $\mu_T = 4.15 \ldots $ and $\mu_H = \sqrt{ 2 + \sqrt{2}} = 1.85\ldots$ \cite{jensen2004improved, duminil2012connective}%
; in particular, $\mu_T < \mu_S < \mu_H$, and therefore the phase transitions occur at different parameter settings on different lattices. To discuss this precisely, we consider two additional graphs:

\begin{defn}[Hexlandia]
Hexlandia, $H_n$, is a graph of $n \times n$ hexagonal tiles arranged as in \Cref{fig:landias}. Each node has population 1.
\end{defn}

\begin{defn}[Trilandia]
Trilandia, $T_n$, is a graph of $n \times n$ triangular tiles arranged as in $\Cref{fig:landias}$. Each node has population $1$.
\end{defn}

Connected 2-partitions on Hexlandia correspond to self-avoiding walks on a dual triangular lattice graph, and connected 2-partitions on Trilandia correspond to self-avoiding walks on a dual hexagonal lattice graph. In particular, $1/\mu_H$ is the phase transition for $\beta_{\lambda}$ in Trilandia, and $1/\mu_T$ is the phase transition for $\beta_{\lambda}$ in Hexlandia. In general, the behavior of Markov chains on connected partitions can depend on which side of the phase transition one is on, and therefore the behavior of the same algorithm can vary widely based on whether the state is discretized into triangles, squares or hexagons. We return to the role that topology the state dual graph plays in \Cref{sec:role_of_discretization}, but for the remainder of this section we focus on gridlandia.

\paragraph*{Summary of experiments.}

The remainder of this section consists of experiments investigating the mixing time and qualitative properties of the flip walk Markov chain in each of three fugacity regimes. Specifically:

\begin{itemize}
    \item In \Cref{sec:metastableslopes}, we explore the mixing time of the Markov chain by measuring statistics about the slope of the partition boundary. 
    \item In \Cref{sec:SLEtests}, we follow up on the apparent success of the critical regime flip walk chain by testing it against known statistics of Schramm-Loewner evolution.
    \item In \Cref{sec:population_balance_rejection_sampling}, we examine the distribution of population balance in the three fugacity regimes.
\end{itemize}
Each experiment computes statistics from an MCMC run through connected partitions on gridlandia. %
In \Cref{sec:counting}, \Cref{sec:thestatisticswemeasure}, and \Cref{sec:proposal_scores}, we summarize precisely how we measure the length of the Markov chain experiments, the summary statistics we use and the precise specifications of the Markov chains, respectively. In \Cref{sec:discussion}, we conclude with a conjecture summarizing the experimental data and related questions.

\paragraph*{Github repository}

Code for reproducing the experiments in this paper is available at: \url{https://github.com/LorenzoNajt/Code_For_Empirical-Sampling-of-Connected-Graph-Partitions-for-Redistricting}

\subsection{Experimental details: Measuring run time}\label{sec:counting}

We use the Metropolis--Hastings algorithm on a reversible Markov chain $M$ that can be represented as an undirected, edge-weighted graph $G_M$. In our case, $G_M$ has many self loops, and instead of simulating those self loops during implementation we can simulate the random number of self-loops taken before moving to a different state. To add to the confusion, we do not simulate every self loop this way, but only a large fraction of them. %
In this section we include definitions for \emph{compressed steps}, \emph{steps} and \emph{accepted proposals} that are intended to clear up the potential ambiguity that can occur when reporting on Markov chains that simulate self-loops. We also specify the computing resources used.

\paragraph*{Counting the number of steps.}%

Recall that the flip walk on $P_2(G)$ picks a node $v \in V(G)$ uniformly at random and a color $c \in \{0,1\}$ uniformly at random, and proposes changing the color of $v$ to $c$. The flip walk accepts this proposal if the coloring remains connected and if the number of colors remains equal to $2$. At a partition $P$, the only nodes that can change color are those adjacent to an edge of $\cut(P)$; call that set $\bdnodes(P)$. This means that instead of uniformly sampling from vertices $v$ until we obtain a sample from $\bdnodes(P)$, we can sample the number of times we expect to draw a sample outside of $\bdnodes(P)$ and then uniformly sample from $\bdnodes(p)$. This is achieved by the following observation:

\begin{observation}
Let $B \subseteq N$ be two finite sets. Let $X_1, X_2, \ldots$ be iid samples from $N$. Let $T$ be the first time that $X_i \in B$. Then $T$ is geometrically distributed according with success probability $|B|/|N|$, and $X_T$ is a uniform sample from $B$.
\end{observation}

Thus, instead of picking a uniform $v \in V$, one can pick $v$ uniformly from $\bdnodes(P)$ and use a geometric random variable with success probability $| \bdnodes(P) | / |V| $ to count the number of self loops taken because of drawing $v \not \in \bdnodes(P)$.

We let \emph{compressed steps} refer to the number of times we draw from $\bdnodes(P)$, and \emph{steps} refer to the total sum of the aforementioned geometric random variables. Finally, if we use an additional score function/Metropolis--Hastings step on top of the flip walk proposal function, we let \emph{accepted proposals} refer to the number of proposals accepted by the Metropolis--Hastings algorithm, and \emph{proposals} refer to the number of proposals made.

\begin{remark}
Only certain nodes in $\bdnodes(P)$ result in legal flips, so that in fact even more of the self-loops of the underlying directed graph could be ignored and wrapped into a geometric random variable. However, we did not do so in the implementation.
\end{remark}

\paragraph*{Measuring the computation time.}

\emph{Computational time} refers to the number of seconds the the 0.2.12 release of the open-source GerryChain software ran for.
All experiments about mixing times and phase transitions on gridlandia ran on a 
Xeon(R) CPU E5-2683 v4 at 2.10Ghz. The annealing experiments were run on an Ubuntu 16.04 machine with 64 GB memory and an Intel Xeon Gold 6136 CPU (3.00GHz) processor.

\subsection{Experimental details: Statistics measured}\label{sec:thestatisticswemeasure}

This subsection will summarize the statistics we use to empirically analyze the behavior of these Markov chains. 

\paragraph*{Boundary slope.}

A natural qualitative measurement of the shape of a simply connected, connected 2-partition of gridlandia is whether it is like a vertical, diagonal or horizontal partition.  The line connecting the endpoints of the boundary of a simply connected partition gives us a simple way to turn that intuition into a quantitative measurement whose long-term behavior can be visualized and measured. This measurement will give us insight into bottlenecks, or metastable regions, in the flip walk.

\begin{figure}
    \centering
    \begin{tabular}{cc}
           \includegraphics[scale = 1]{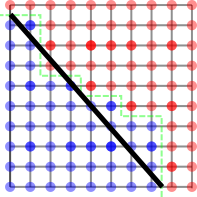} & \includegraphics[scale = .3]{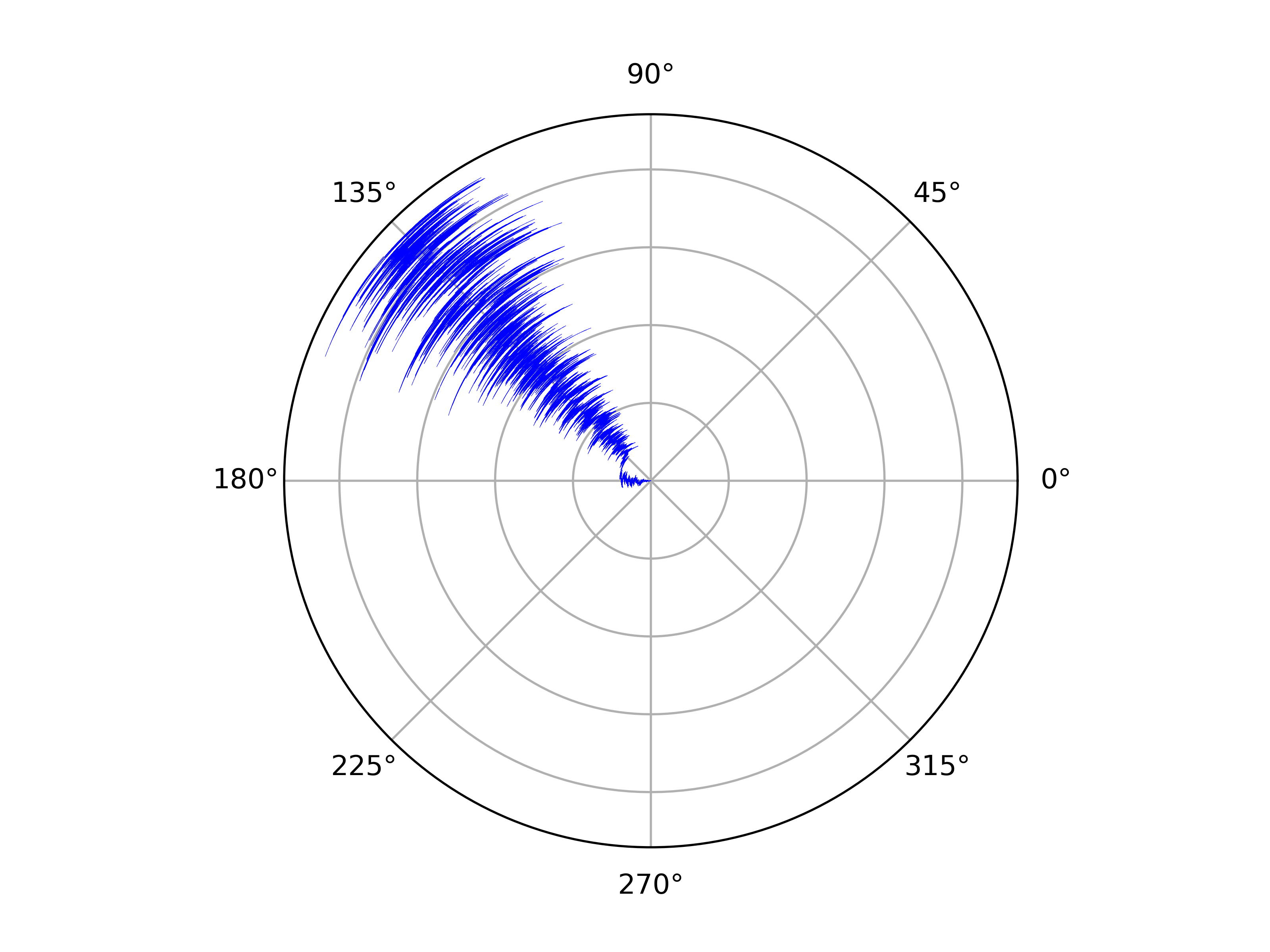}
    \end{tabular}

    \caption{(Left) The boundary line segment. (Right) An example boundary slope plot.}
    \label{fig:boundaryslopeplot}
\end{figure}

\begin{defn}[Boundary line segment]
Let $P$ be a connected partition of gridlandia $G_n$. Then $P$ induces a connected $2$-partition $Q$ of the boundary of $G_n$, which is a cycle with $4n - 4$ nodes. If $\cut(Q)$ is non-empty, then it has 2 edges, $e_1$ and $e_2$, and the boundary line segment is the line segment connecting the midpoints of $e_1$ and $e_2$. If $Q$ is the trivial partition, and $\cut(Q) = \emptyset$, then we define the boundary line segment to be $\bot$.
\end{defn}

We summarize a time series of boundary line segments via their slopes, which we plot in polar coordinates with the radial axis denoting time. %
That is, 
\begin{defn}[Boundary Slope Plot]
 Let $P_1, \ldots, P_n$ be a sequence in $P_2(G_n)$, and let $l_1, \ldots, l_n$ be the corresponding boundary slopes. For each $l_i \not = \bot $, there are two associated unit vector directions $\pm v_i$. We choose signs so that $v_i$ is closer to $v_{i-1}$ than $-v_{i-1}$ and draw an intervals connecting $(i-1)v_{i-1}, iv_i$; in the limit, this amounts to picking a continuous path of unit vectors that are aligned with the boundary slope. %
\end{defn}

Our implementation %
produces the sequence $(P_1, \ldots, P_n)$ by taking every $10,000$ steps (which include self loops, as discussed in \Cref{sec:counting}); %
this choice generally captures every compressed step. For aesthetics, we interpolate between points $(\theta, r)$ and $(\theta', r')$ in polar coordinates with the curved line $( \lambda \theta + (1 - \lambda) \theta', \lambda r + (1 - \lambda) r')$. The radial direction measures time, and the stretching that comes with being further from the origin is a distortion that has nothing to do with the underlying path of partitions; only the angle represents a feature of the corresponding partition. Finally, we draw circles at regular intervals to give a sense of time segments. 
An example boundary line segment and boundary slope plot is given in \Cref{fig:boundaryslopeplot}.

\paragraph*{Flips and edge cuts.} %

Here, we summarize two closely related statistics that can be used to see metastable regions in the Markov chain. Unlike measuring the boundary slope, these store an aggregate measure of variation, rather than a detailed history. This can make it harder to see metastability after sufficiently long runs, but these statistics are especially useful because they generalize easily to $k$-partitions. Additionally, patterns that also have to do with the internal behavior of partition boundaries, such as those we highlight in \Cref{sec:role_of_discretization}, are easier to see from these plots since boundary slope plots only record the relative behavior of the endpoints of the partition boundary.

\begin{defn}[Flips]
We define a function $\textrm{flips} : V \to \mathbb{N}$ by setting $\textrm{flips}(v)$ to be the number of times the district containing $v$ changed: $\textrm{flips}(v) = | \{ i \in \{1, \ldots, n-1 \} : f_i(v) \not = f_{i+1}(v) \} |$. Usually we will plot $\textrm{lflips} := \log( \textrm{flips} + 1)$ instead.
\end{defn}

We plot $\textrm{lflips}$ using a heat map where red means an node is flipped more often, and blue means it is flipped less often. 

\begin{defn}[Edge Cuts]
We define a function $ c : E \to \mathbb{N}$ by setting $c(e)$ to be the number of times that $e$ was a cut edge of a partition $P_i$: $c(e) := | \{ i \in \{1, \ldots, n\} : e \in \cut(P_i) \}|$.
\end{defn}

We visualize cut edges as heat maps on the state dual graph; lighter green/yellow edges are those that are cut more than darker green/purple edges. Edge cuts are a useful statistic for examining proposals that do not flip individual vertices, such as the tree based proposals. See \Cref{fig:metamandering_images} for an example.

\subsection{Experimental details: Proposal \& score function}\label{sec:proposal_scores}

Here we provide further details of a flip walk that is modified to reject proposals leading to heavily population unbalanced states, and that is further modified by using the Metropolis--Hastings algorithm with the score function $\lambda^{\cut}$.

\paragraph*{Hard population constraints, rejection  sampling population constraints.}

We define the population balance score $\text{PopBal}$ of a partition $(A,B)$ of $n \times n$ gridlandia as $\text{PopBal} ( A,B) = \max ( |A|/\frac{n}{2}, |B|/\frac{n}{2} )$. We set a threshold $\alpha > 0$, and define the flip walk with population constraint $\alpha$, $\text{Flip}_{\alpha}$ to be the flip walk modified to reject all proposals that would produce a partition with $\text{PopBal(A,B)} > 1 + \alpha$. %

\paragraph*{Metropolis--Hastings for $\lambda^{\cut}$ compactness constraint.}

Fix some $\lambda > 0$. We use the Metropolis--Hastings algorithm on top of $\text{Flip}_{\alpha}$ with score function $f( (A,B)) = \lambda^{\cut(A,B)}$, where $\cut(A,B) = \{ e \in E(G) : e = \{v,w\}, (v \in A \wedge w \in B) \vee ( v \in B \wedge w \in B) \}$. We denote the resulting Markov chain by $\text{Flip}_{\alpha, \lambda}$.

 \subsection{Experiment: Metastable Regions in the flip walk chain}\label{sec:metastableslopes}
 
In this section, we will see that the boundary slope plots give insights into bottlenecks in $\text{Flip}_{\alpha, \lambda}$. We break our analysis down into the three phase transition regimes of the $\lambda^{\cut}$ score function.

 \paragraph*{The subcritical case.} %

Subcritical self-avoiding walks in the grid are known to converge to geodesics \cite{duminil2014supercritical}. In the context of connected partitions, this means that the score function $\beta_{\lambda}$ strongly prefers minimizing the boundary length of a partition. Unlike the self-avoiding walk case, however, in redistricting we have to contend with population balance. It turns out that the behavior of the subcritical cases depends strongly on whether or not the two districts are required to have balanced population.

\begin{figure}
    \centering
     \begin{tabular}{c@{\hspace{-.3in}}c}
    \includegraphics[width = 200pt]{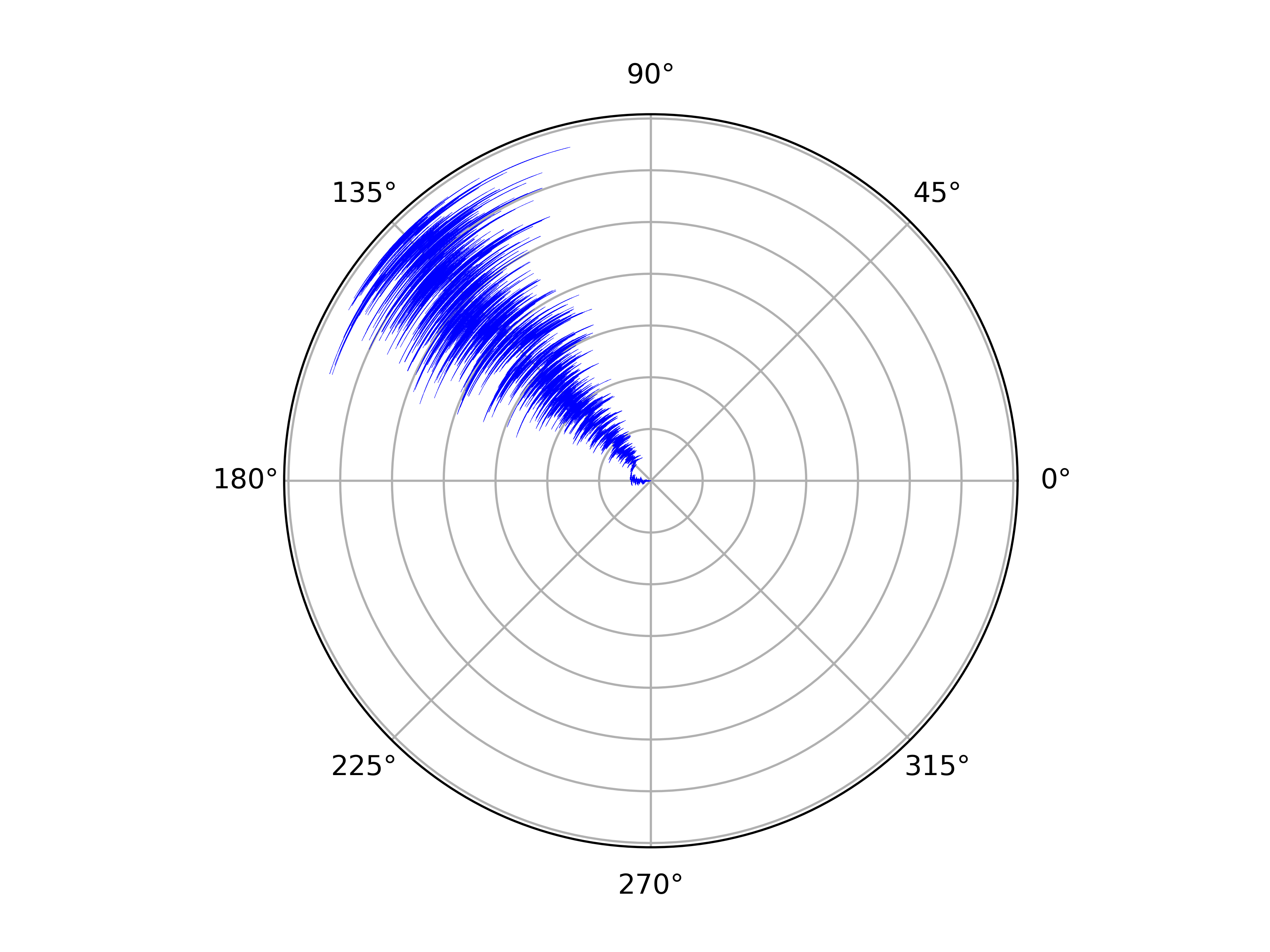}
 & \includegraphics[scale = .1]{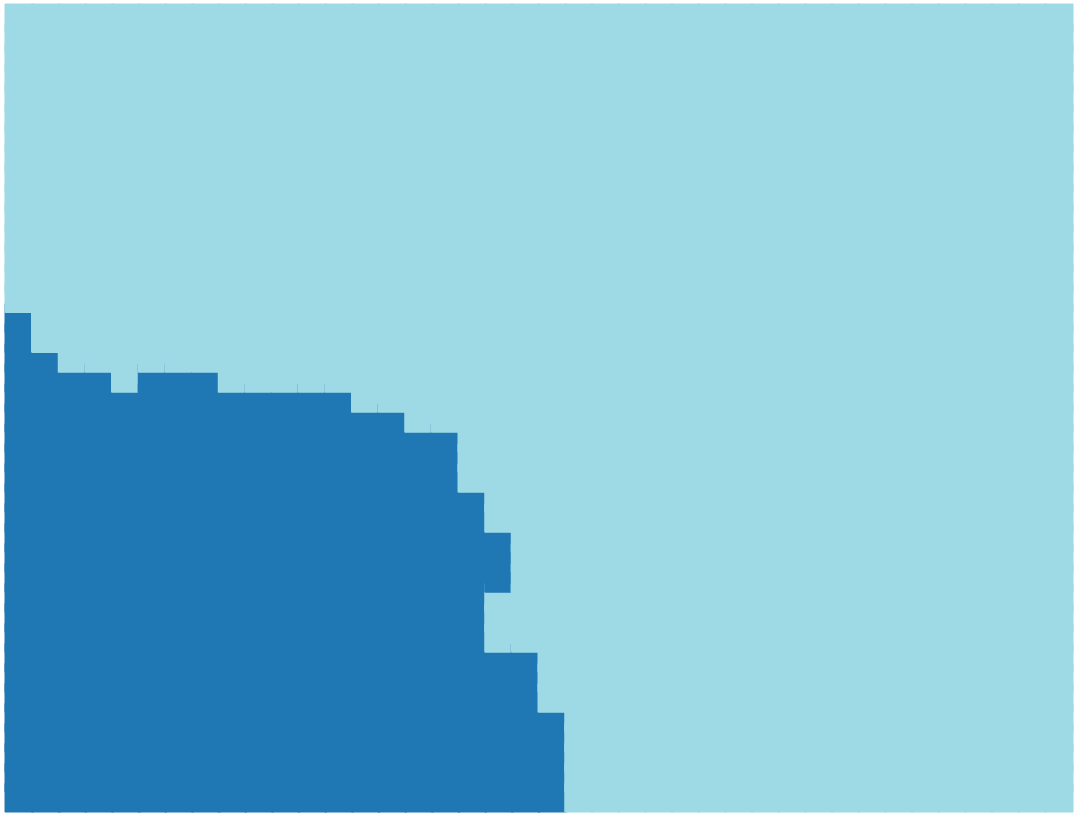}\\ 
 
  \includegraphics[width = 200pt]{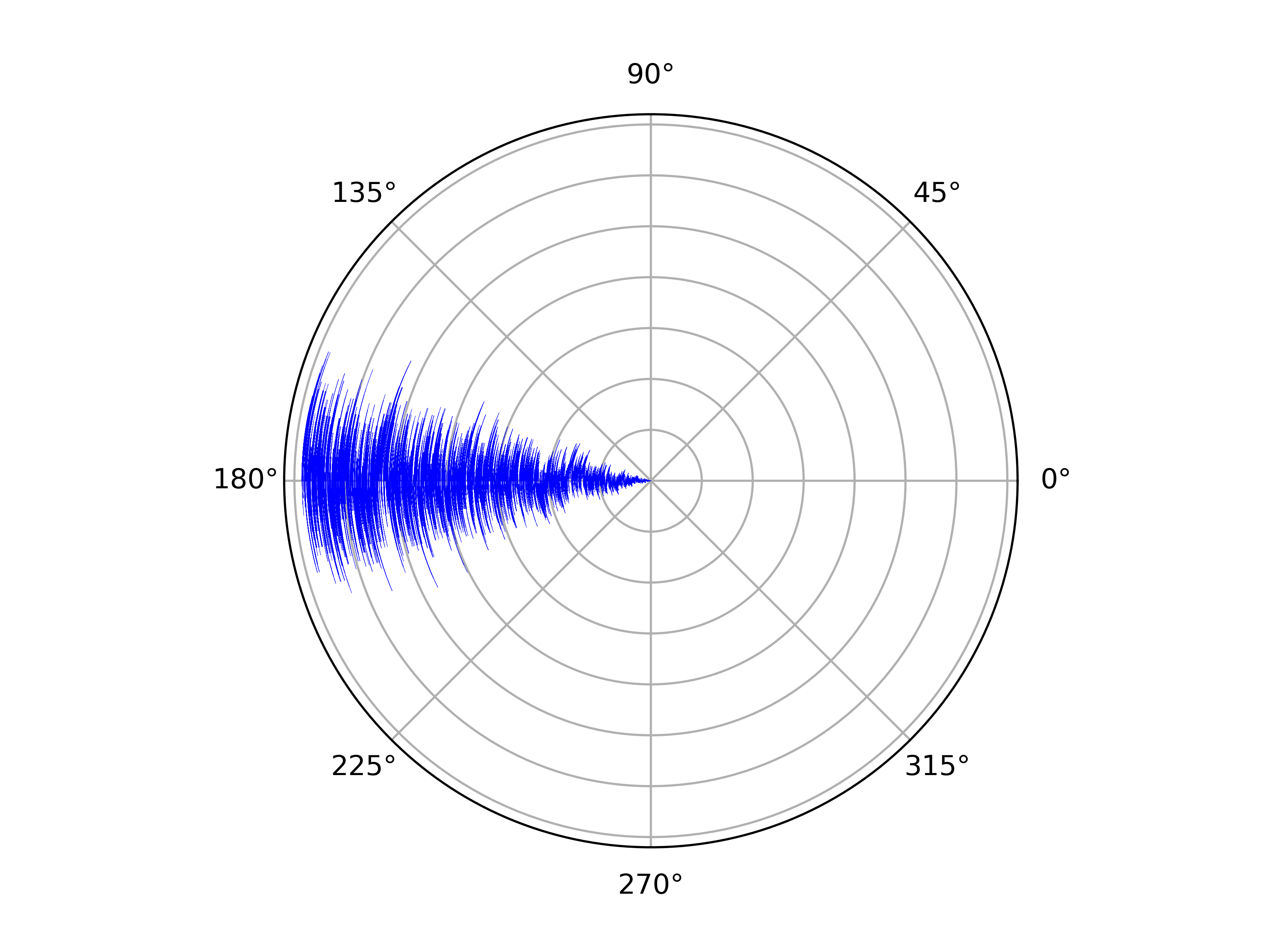}
 & \includegraphics[scale = .1]{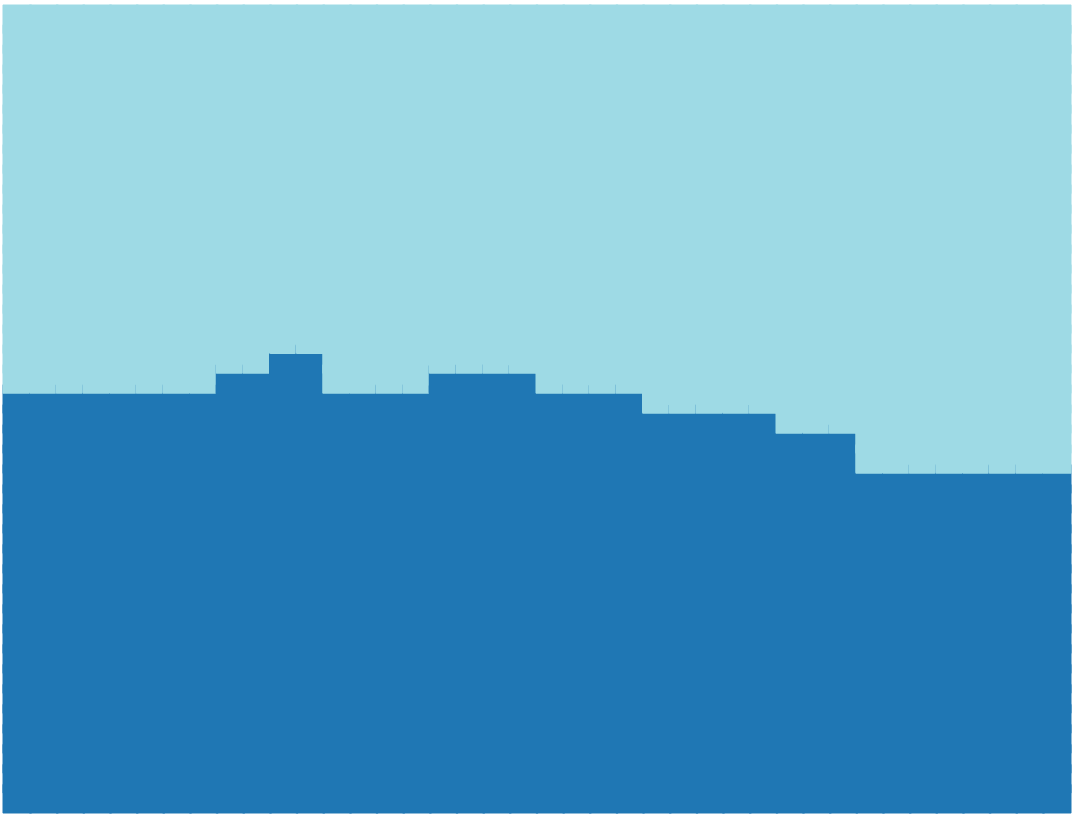}
    \end{tabular}

    \caption{(left) Boundary slope plot and (right) the end state, from a subcritical ($\lambda = \frac{1}{2\mu} = .189\ldots$) flip walk constrained to population balance 10 percent (bottom) and 50 percent (top). The top plot was run for $21,000,000$ compressed steps, which amounted to $703,328,732,936$ proposals, over roughly one month of computing time. The bottom plot was run for $23,000,000$ compressed steps, $712,097,053,677$ proposals, over the same time period.
    }
    \label{fig:boundary_slope_subcritical}
\end{figure}

If the population balance is unrestricted, the chain appears to mix thoroughly in its state space, which consists of partitions where one block occupies most of the region and the other block is a small bubble. As the population constraint is tightened, the ability of the Markov chain to move around the state space becomes more tightly constrained. We illustrate the evolution of 10\% and 50\% population balances in \Cref{fig:boundary_slope_subcritical}: with strong population constraints, the chain gets stuck around one of the partitions satisfying the population constraint with the shortest possible boundary. We note that the results of these experiments change when $\lambda$ is closer to the critical fugacity, and will discuss that case in more detail after reviewing the critical and supercritical regimes. %

In \Cref{sec:population_balance_rejection_sampling}, we revisit the influence of population balance on subcritical sampling through rejection sampling.

\begin{figure}
    \centering
    \includegraphics[scale = .4]{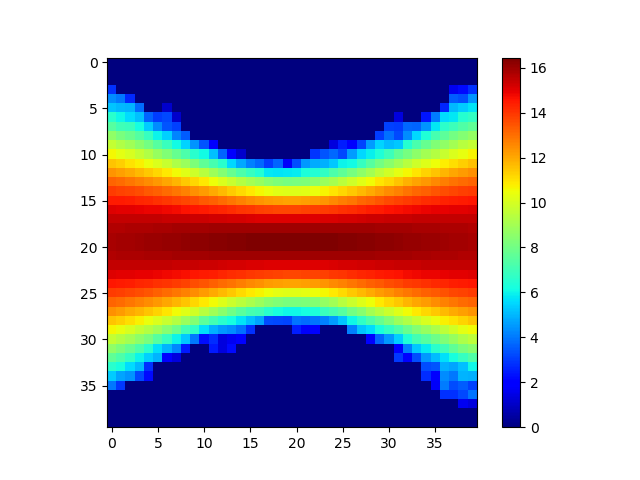}
    \caption{$\log( \text{flips} + 1)$, after 254 hours (equivalently, 64,054,464,437,342 proposals) on a $40 \times 40$ grid graph. pop bounds 5 percent. This is stuck in a bottleneck set; it has not reached the equally probably region around the vertical partition. Loosening either the population constraint or bringing the fugacity closer to the critical value can help mixing.}
    \label{fig:lflips_plot_subcritical}
\end{figure}

\paragraph*{The supercritical case.} %

\begin{figure}
    \centering
    \includegraphics[height = 2cm, width = 2cm]{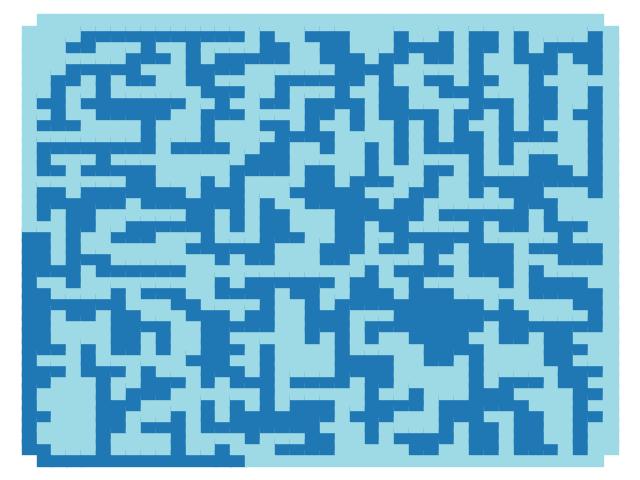}
    \caption{A typical supercritical sample.}
    \label{fig:supercritical_sample}
\end{figure}

\begin{figure}
    \centering
    \includegraphics[scale = .4]{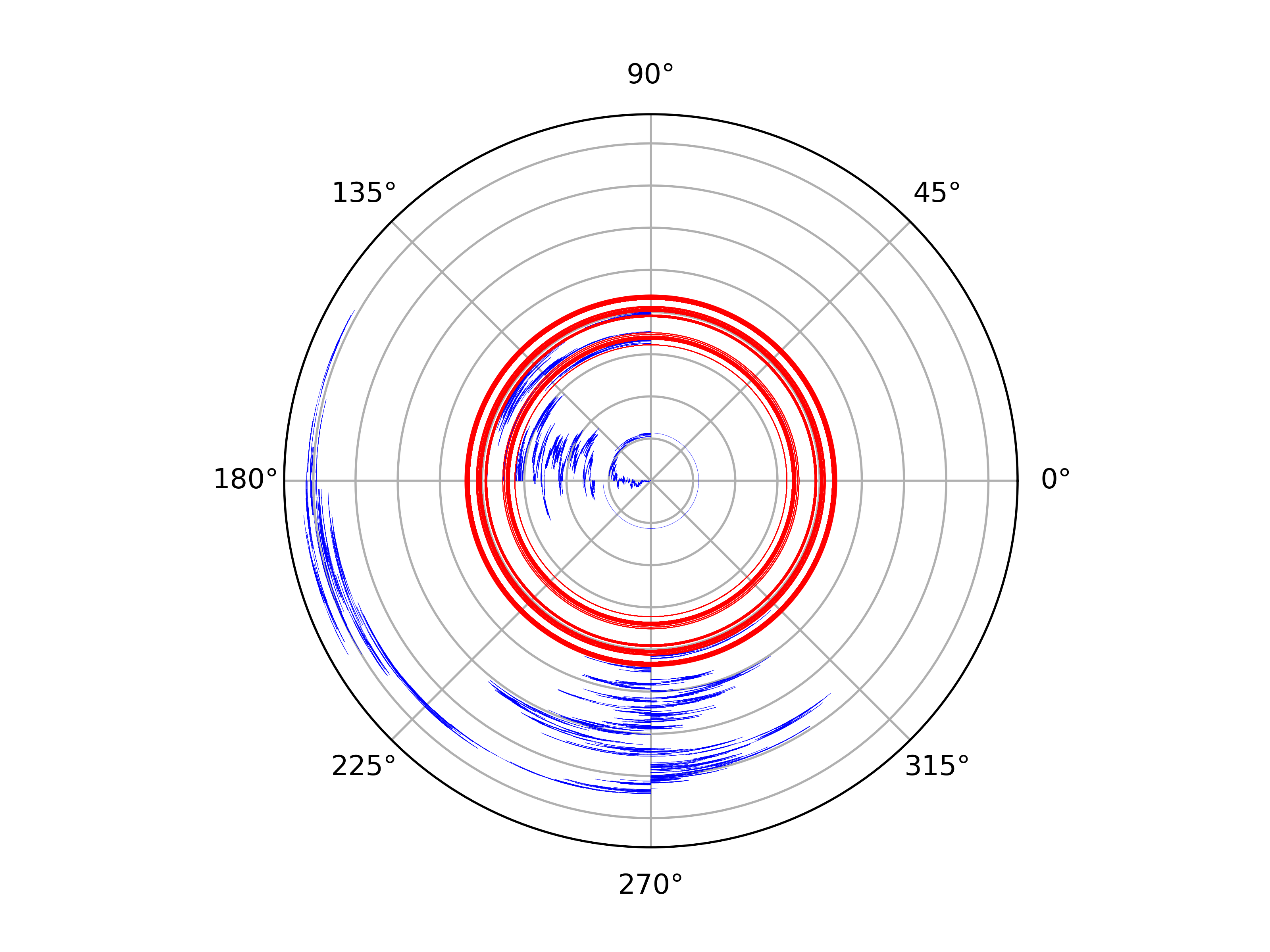} 
    \caption{Boundary slope plots (left) for a super critical $(\lambda = 1)$ run with population constraint $100\%$. The apparent rapid rotation in the beginning of the run was due to the boundary end points ending up near the two corners of one edge. This was run for $11,000,000$ compressed steps, or $20,302,926,563$ proposals, over the course of one month of computing time.}
    \label{fig:mixing_supercritical}
\end{figure}

The supercritical case is generally unsuitable for redistricting, since the boundaries are almost-surely space filling \cite{duminil2014supercritical}; see \Cref{fig:supercritical_sample} for a typical example. %
Experiments show that the flip walk Markov chain also has bottlenecks that are revealed by the boundary slope statistics, see \Cref{fig:mixing_supercritical}.  %
On the other hand, the supercritical case appears to mix in space, in the sense (see \cite{moore2011nature}) that the configuration of the partition boundary in the interior of gridlandia seems to be independent of the partition restricted to the boundary of gridlandia.
 
 \paragraph*{The critical fugacity case.}\label{sec:critical}
 
\begin{figure}
     \centering
     \begin{tabular}{cc}
             \includegraphics[scale = .3]{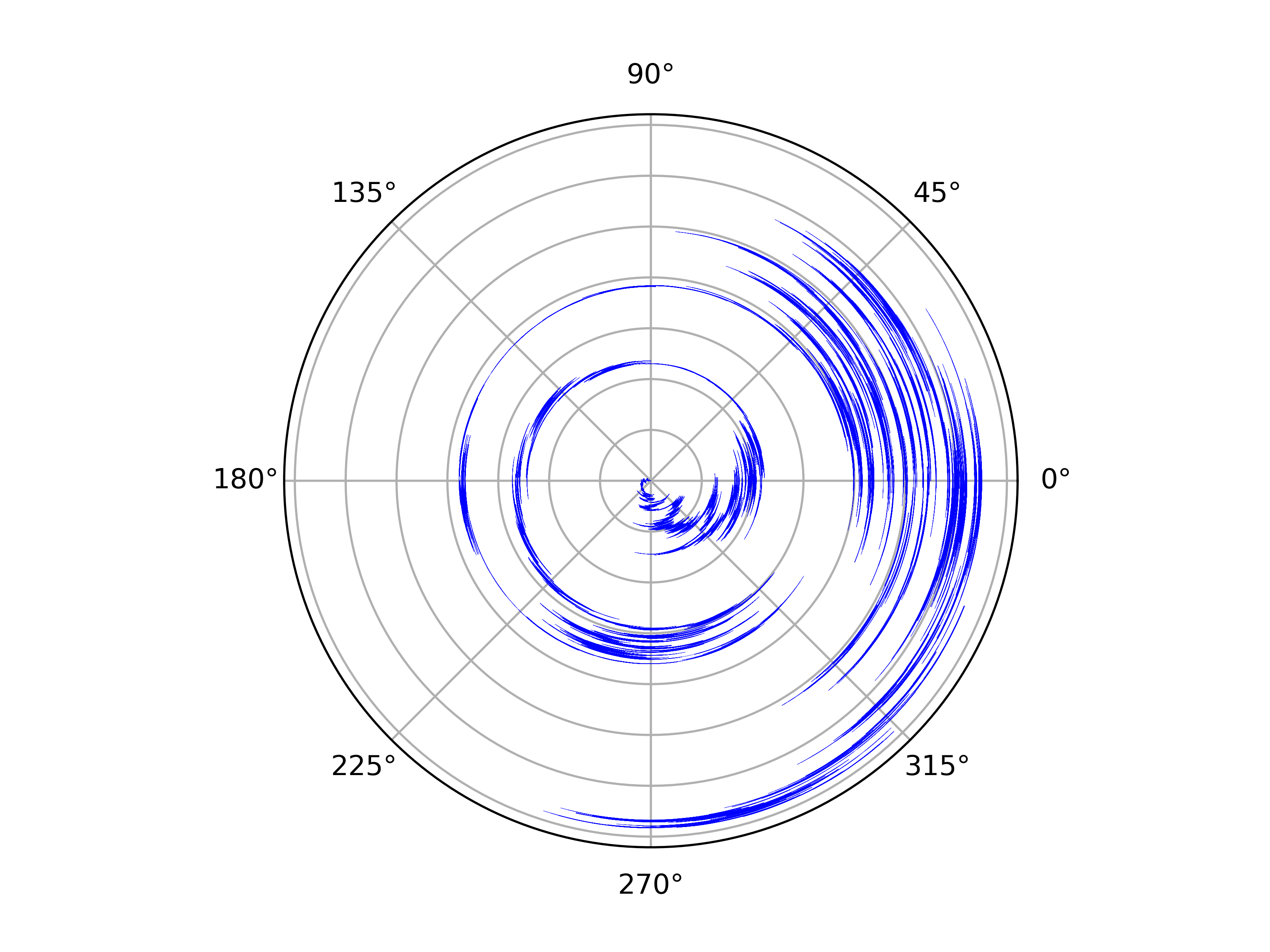}  &  \includegraphics[scale = .1]{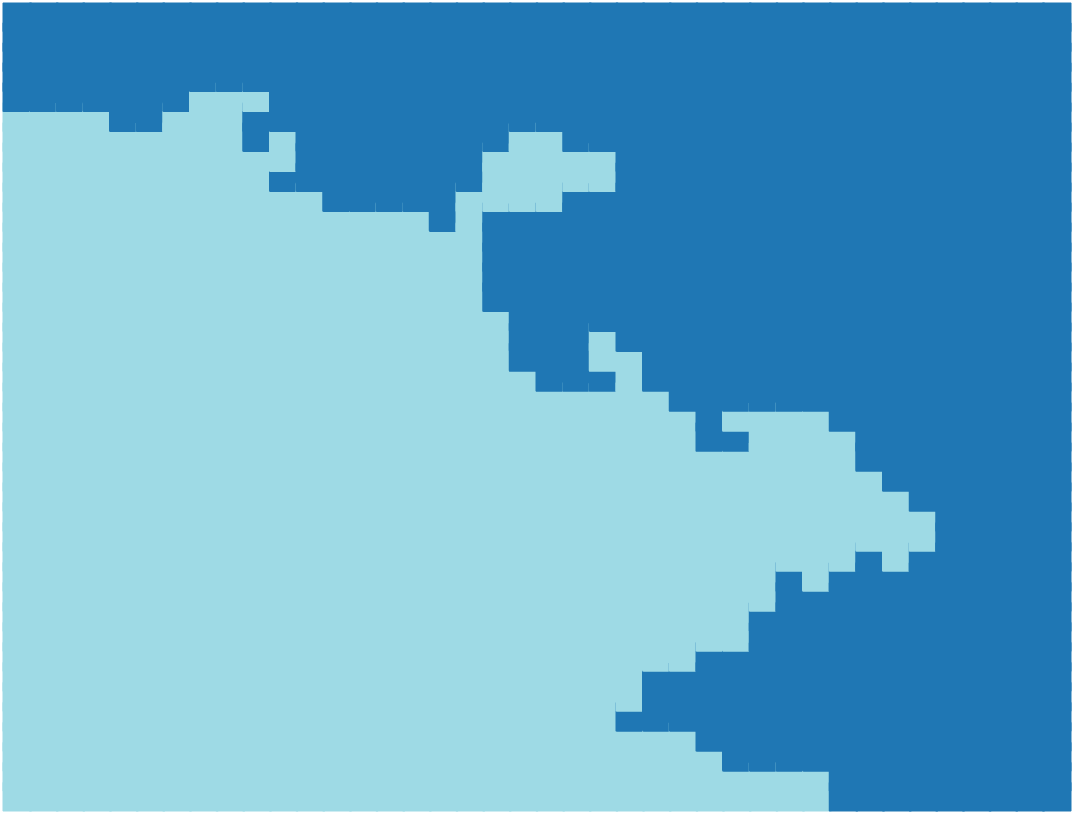}
     \end{tabular}

     \caption{Boundary Slope Plots for $\lambda = 1/\mu$ (critical regime), with population constraints $10 \%$, along with the final state on a $40 \times 40$ grid graph. Run for $25,000,000$ compressed steps, amounting to $354,337,192,915$ proposals, over the course of approximately one month of computing time. The metastable region that was apparent in the subcritical case is not an obstruction here. }
     \label{fig:critical_fugacity_mixing}
 \end{figure}

 Unlike the previous two cases, this setting of the score function does not cause the chain to get trapped in a metastable region corresponding to the slope of the boundary. This fact can be seen in the boundary slope plot \Cref{fig:critical_fugacity_mixing}. Additionally, the chain appears to continue to mix rapidly even with strong population constraints.
Still, whether or not the flip walk is rapidly mixing at this setting of $\lambda$ remains unknown, since there could be more subtle bottlenecks. 

To explore that possibility, more intricate tests of mixing are needed; in the next section we show that samples from the flip walk at $\lambda = \nicefrac{1}{\mu}$ produce empirical statistics that agree with predictions from Schramm--Loewner evolution. This is further, although still inconclusive, evidence of rapid mixing of the flip walk on gridlandia at critical fugacity.

\paragraph*{When does the phase transition occur?}

\begin{figure}
    \centering
    \begin{tabular}{c@{\hspace{-.2in}}c@{\hspace{-.2in}}c}
           \includegraphics[scale = .2]{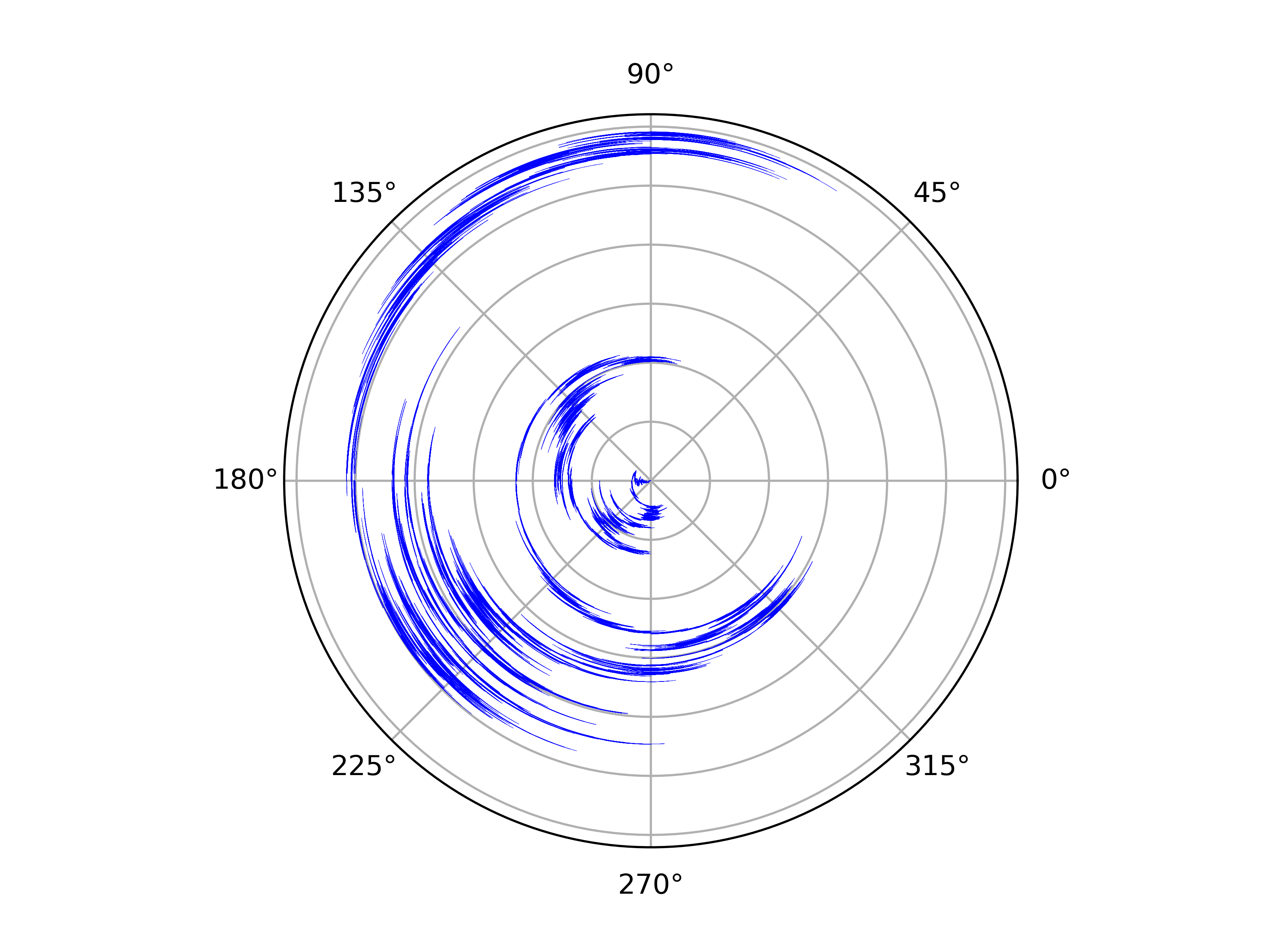}  &  \includegraphics[scale = .2]{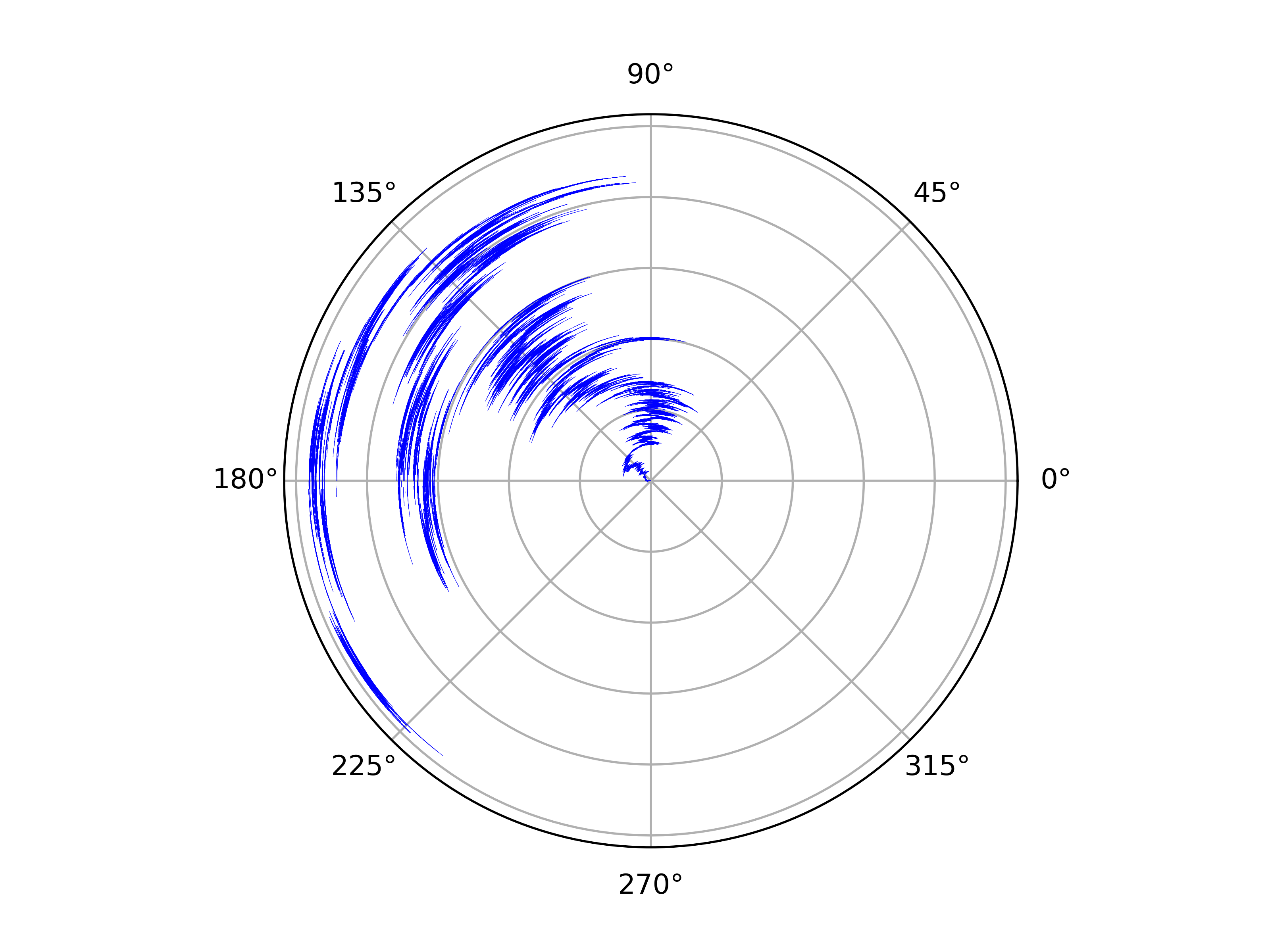} & \includegraphics[scale = .2]{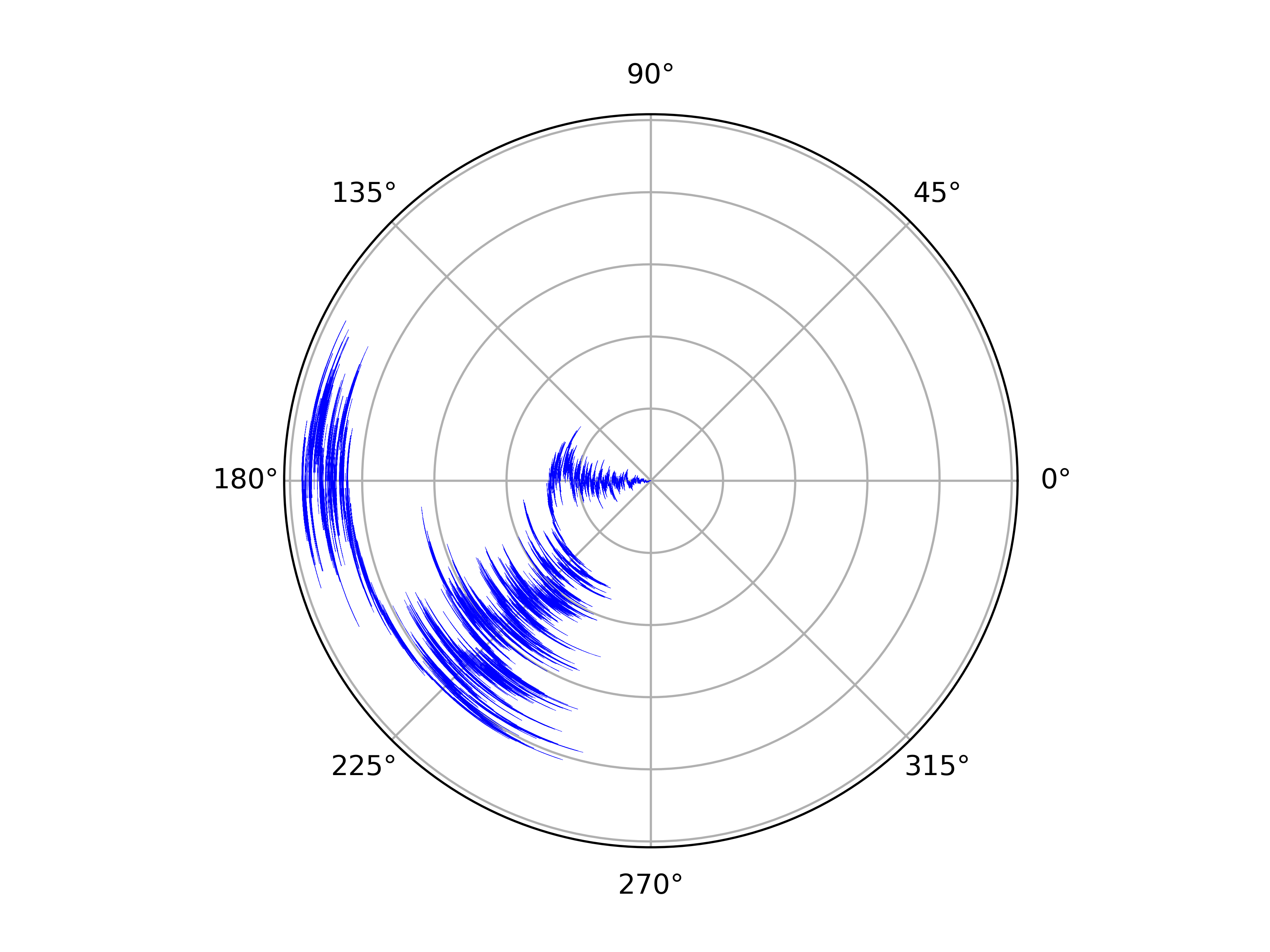}
    \end{tabular}

    \caption{ (Left) Setting $\lambda = 1/\mu - .3 = .35\ldots$, population bound at $50 \%$ and using a $40 \times 40$ grid graph results in a run that looks more like the critical case than the subcritical case. (Middle) The same but $\lambda = .32$, (Right) same but $\lambda = .29$. Each of these was run for about a week of computing time, which amounted to the order of 250,000,000,000 proposals each.}
    \label{fig:variable_lambda_subcritical}
\end{figure}

The phase transition results in \cite{duminil2014supercritical} are asymptotic, and for any given grid graph the distinctions between the three regimes are fuzzy. For instance, in \Cref{fig:variable_lambda_subcritical} we take $\lambda$ at three values closer to $\nicefrac{1}{\mu}$ than $\nicefrac{1}{2\mu}$, and see that doing so produces behavior that is more the critical regime than the subcritical regime. In particular, the bottlenecks we saw in the subcritical regime become gradually less pronounced as $\lambda$ approaches $\nicefrac{1}{\mu}$. Exactly how the asymptotic phase transitions kick in seems like a challenging question, and one with implications for sampling maps for redistricting.
 \subsection{Experiment: SLE based test of mixing at criticality}\label{sec:SLEtests}

Although the boundary slope plots suggest bottlenecks in both the subcritical and supercritical regimes, they do not suggest the same bottlenecks in the critical regime. In this section, we provide another test for mixing at the critical regime, and show that the Markov chain passes this test.

\paragraph*{Background on Schramm--Loewner Evolution.}
The theory of Schramm--Loewner evolution (SLE) can be used to make a prediction about the expectation of a statistic on the Markov chain of connected 2-partitions in the critical regime. We thus obtain a test for mixing by empirically calculating this expectation and comparing it to the theoretically predicted value.  Our experiments in this section are indebted to the investigation in \cite{kennedy2002monte}, which used the same statistic to verify conjectures about scaling limits of self-avoiding walks. 

Schramm--Loewner evolution concerns a family of probability distributions, $\SLE_{\kappa}$ with $\kappa \in [0,\infty)$, on compact subsets of the upper half plane $\mathbb{H}$. For a precise definition of Schramm--Loewner evolution, see \cite{berestycki2014lectures}.  Many discrete models from statistical physics converge to $\SLE_{\kappa}$ for appropriate values of $\kappa$. In particular, $\SLE_{8/3}$ has close, albeit conjectural, relationships with self-avoiding walks. 

$SLE_{8/3}$ is a probability distribution over simple paths from $0$ to $\infty$ in $\mathbb{H}$. This distribution is invariant under scaling, so if $\gamma$ is a sample from $SLE_{8/3}$, so is $\alpha \gamma$, for any $\alpha > 0$. In particular, if $a,b \in \partial \mathbb{D}$, the scale invariance of $\SLE_{8/3}$ means that the distribution over simple paths in $\mathbb{D}$ from $a$ to $b$ obtained by conformally mapping $\mathbb{D}$ to $\mathbb{H}$ and sending $a$ to $0$ and $b$ to $\infty$ is well-defined; this distribution over paths in $\mathbb{D}$ is called chordal SLE from $a$ to $b$.

Let $D_n$ denote the nearest neighbor graph on the intersection of $\mathbb{D} = \{ z \in \mathbb{R}^2 : |z| \leq 1 \}$ with $\frac{1}{n} \mathbb{Z} \times \frac{1}{n} \mathbb{Z}$. We recall the conjecture discussed in \cite{duminil2014supercritical, lawler2002scaling} that a $\beta_{1/\mu}$-sample from a self-avoiding walk from $-1$ to $1$ in $D_n$, considered as the trace of a path in the plane, will converge to $\SLE_{8/3}$ in $\mathbb{D}$ from $-1$ to $1$ as $n \to \infty$; a similar conjecture holds for other pairs of points in the boundary. Thus, if a sample from the flip walk closely approximates a $\beta_{1/\mu}$-sample from a self-avoiding walk, then a sample from the flip walk will closely approximate a sample from $\SLE_{8/3}$.  

Explicit formulas for the probabilities that the upper half-plane $\SLE_{8/3}$ hits certain sets are known \cite{kennedy2002monte, lawler2003conformal}, and we use those to design tests of mixing. %
In particular, we use the same statistics as those used by the investigation in \cite{kennedy2002monte}, which relied on the calculation in \cite{lawler2003conformal} of the probability that an $\SLE_{8/3}$ path in the upper half plane $\mathbb{H}$ intersects $B_r(x)$, the ball of radius $r$ centered around $x \in \mathbb{R} \subseteq \mathbb{H}$. We will denote this probability by $\phi(r,x)$. What follows is the description of an experiment using this to test mixing of the flip walk at fugacity $1/\mu$: 

\begin{enumerate}
    \item We are given a sequence of chordal self-avoiding walks $\gamma_1, \ldots, \gamma_N$ as the boundary paths of a sequence of connected $2$-partitions of $D_n$ obtained from a length $M$ run of the flip walk Markov chain at $\lambda = 1/\mu$.
    \item For each $i$, map $\gamma_i$ to $\mathbb{H}$ via a conformal map from $\mathbb{D}$ to $\mathbb{H}$ sending one endpoint of $\gamma$ to $0$ and the other to $\infty$. Call the resulting path $\bar{\gamma}_i$.
    \item Set $\hat{\phi}(x,r) = \frac{1}{N} \sum_{i = 1}^N 1_{ B_r(x) \cap \bar{\gamma}_i \not = \emptyset}$.
\end{enumerate}
 If the chain is rapidly mixing, and ignoring the significant issues of discretization, $\hat{\phi}(x,r)$ should approximate with $\phi(x,r)$. 
 
\begin{figure}
    \centering
    \begin{tabular}{c}
      \includegraphics[scale = .5]{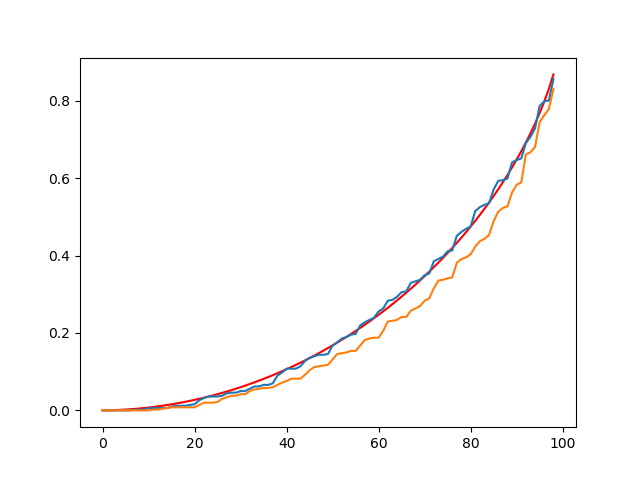}   \\       \includegraphics[scale = .5]{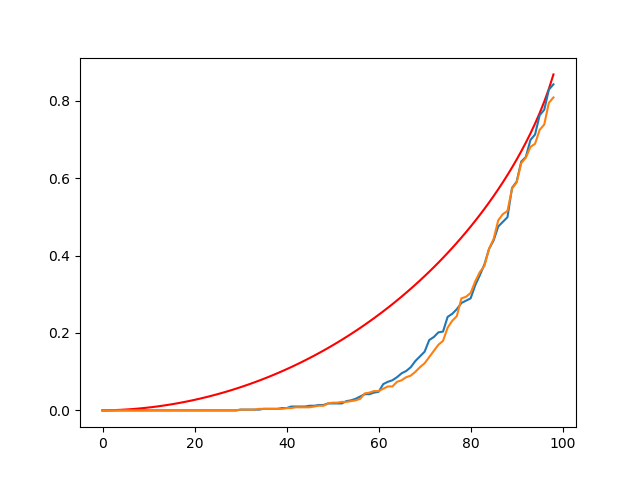} 
    \end{tabular}
    \caption{ %
     Taking $n = 20$ (Top) and $n = 30$ (Bottom), with $\gamma_i$ obtained after $100,000$ steps each with $N = 500$. Comparing $\hat{\phi}(r,x)$ with $\phi(r,x)$ for $x = \pm 1$ and $r \in \{ i / 101 : 0 \leq i \leq 99 \}$. The solid red line is $\phi(1, r) = \phi(-1, r)$, and the other two lines are $\hat{\phi}(1,r)$ and $\hat{ \phi}(-1,r)$. Apparently $100,00$ is sufficient for mixing when $n = 20$, at least for this statistic, but not when $n = 30$.}
    \label{fig:SLE_experiment_fixed_endpoint}
\end{figure}

\paragraph*{SLE experiments.}
 We present two experiments. First, we fix the endpoints at $-1 + 0 i$ and $1 + 0i$, meaning that we start with a connected partition that has a boundary path from $-1$ to $1$, and reject all proposals of the flip walk that would change the endpoints; results are displayed in \Cref{fig:SLE_experiment_fixed_endpoint}. %
Since we do not appear to need many steps to see agreement between the theoretical and empirical statistics, we take this as evidence that the flip walk mixes rapidly at the critical fugacity.

\begin{figure}
    \centering
    \hspace{-.8cm} \begin{tabular}{@{}c@{\hspace{-.15in}}c@{}} \includegraphics[scale = .3]{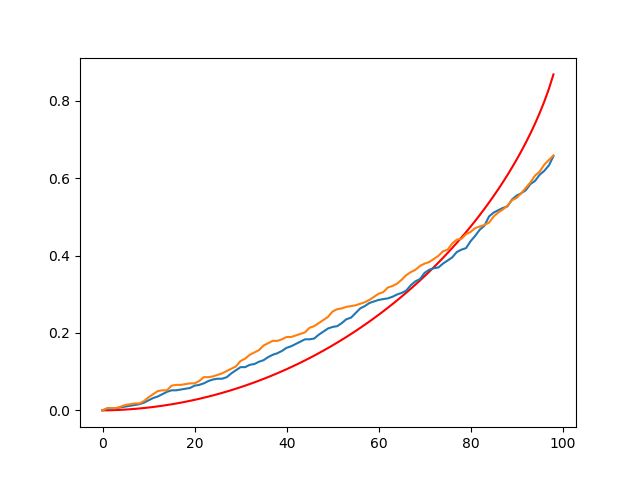}
         &  \includegraphics[scale = .3]{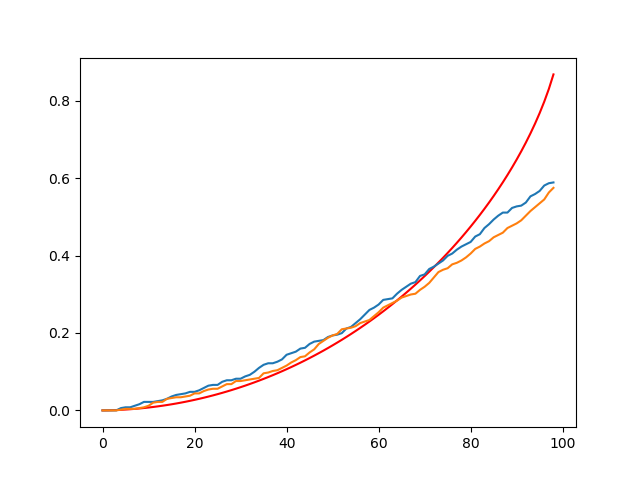}
         
    \end{tabular}

    \caption{With variable endpoints on a $30\!\times\!30$ (left) and $40\!\times\!40$ (right), with $N = 501$ independent runs starting from the horizontal partition for $1,000,000$ steps each. Notation as in \cref{fig:SLE_experiment_fixed_endpoint}. %
    }
    \label{fig:SLE_experiment_variable_endpoint}
\end{figure}

For the second experiment, we drop the fixed-endpoint requirement. In this case, we do not see strong agreement between $\phi(x,r)$ and $\hat{\phi}(x,r)$, as can be seen in \Cref{fig:SLE_experiment_variable_endpoint}. This may have as much to do with the discretization of the disk into $D_n$ as the mixing time of the random walk.

\subsection{Experiment: Population balance. }\label{sec:population_balance_rejection_sampling}

\begin{figure}
     \begin{tabular}{@{}c@{\hspace{-.1in}}c@{\hspace{-.1in}}c@{}}
       \hspace{-.55cm} 
       \includegraphics[scale = .2]{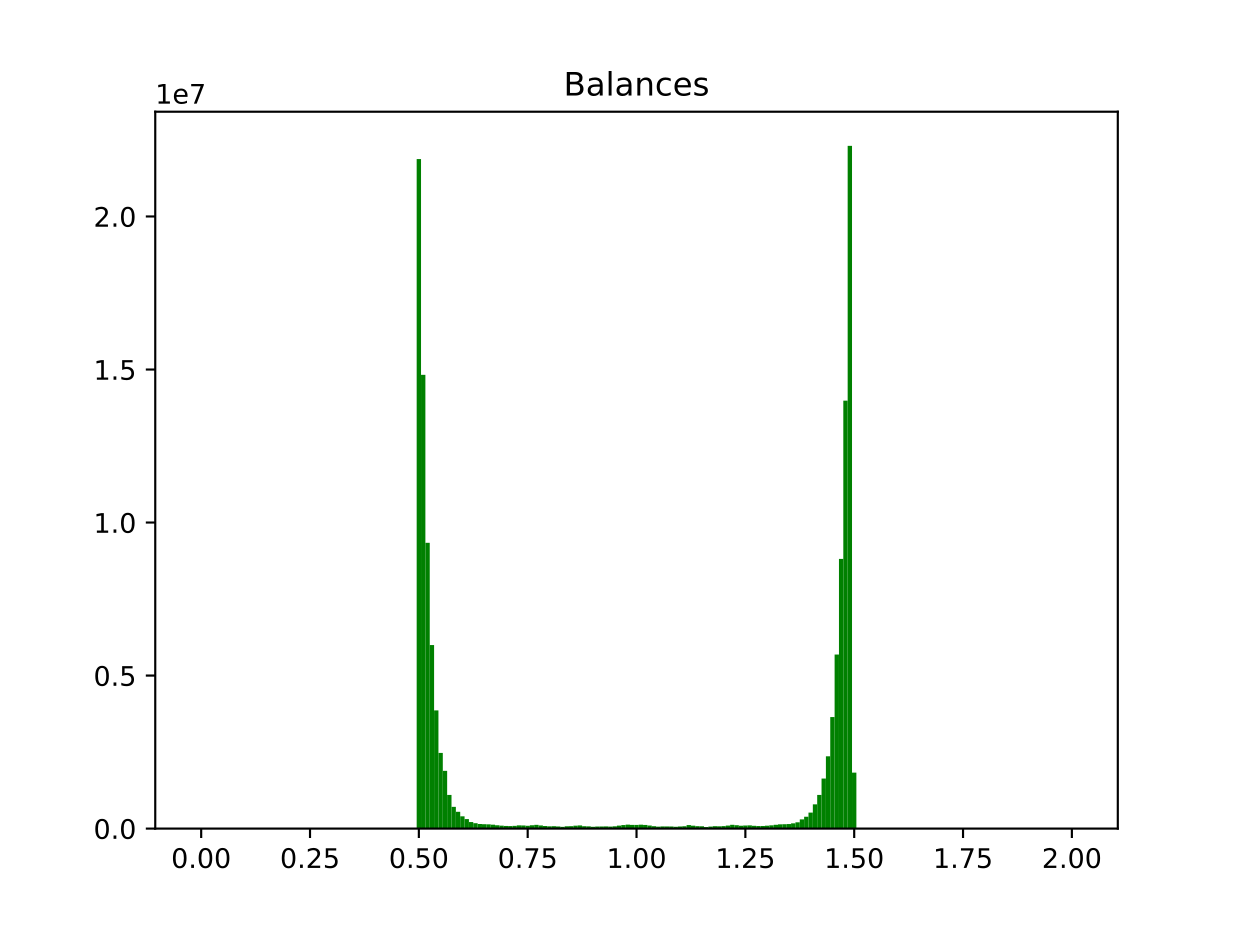}  & \includegraphics[scale = .2]{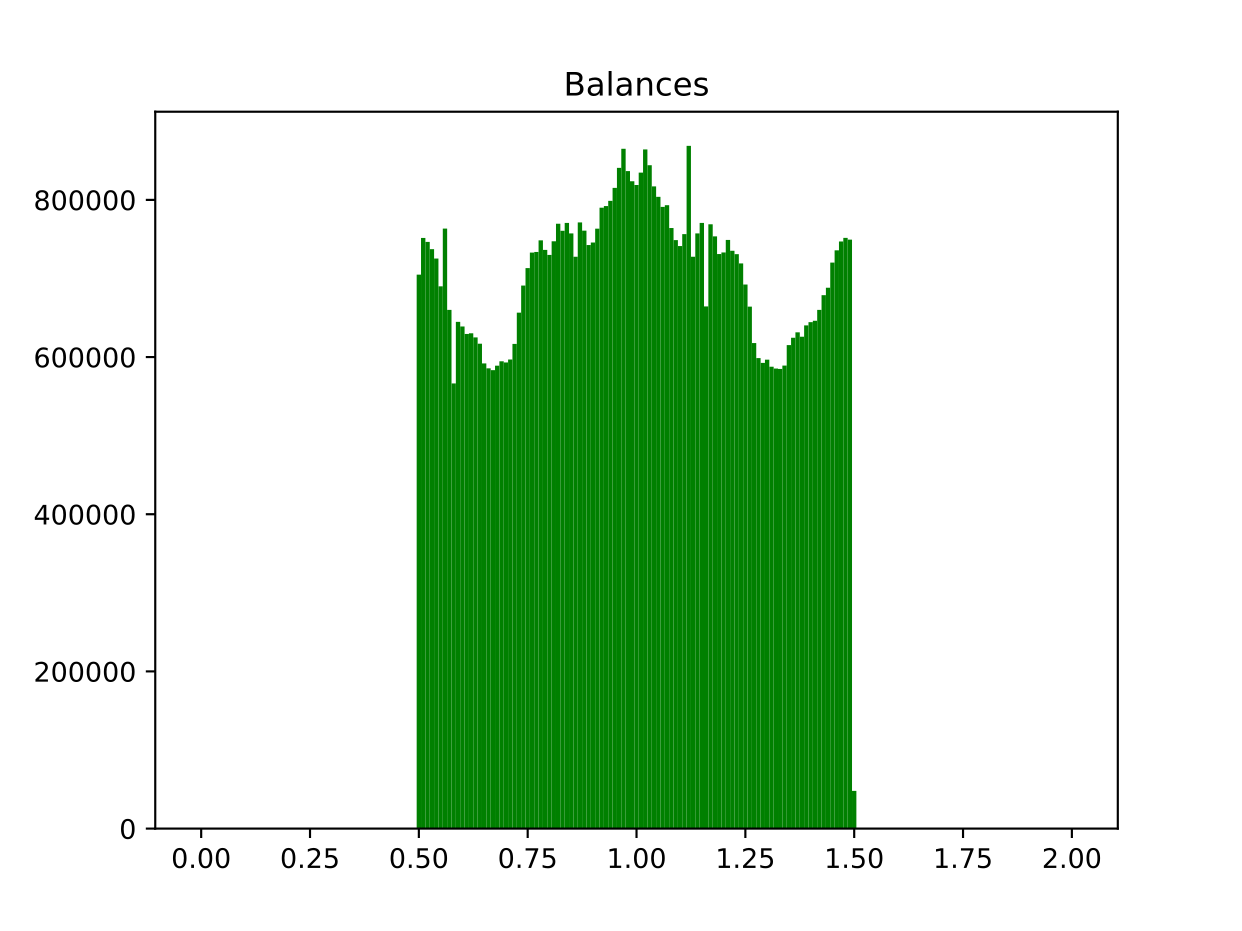} 
         & \includegraphics[scale = .2]{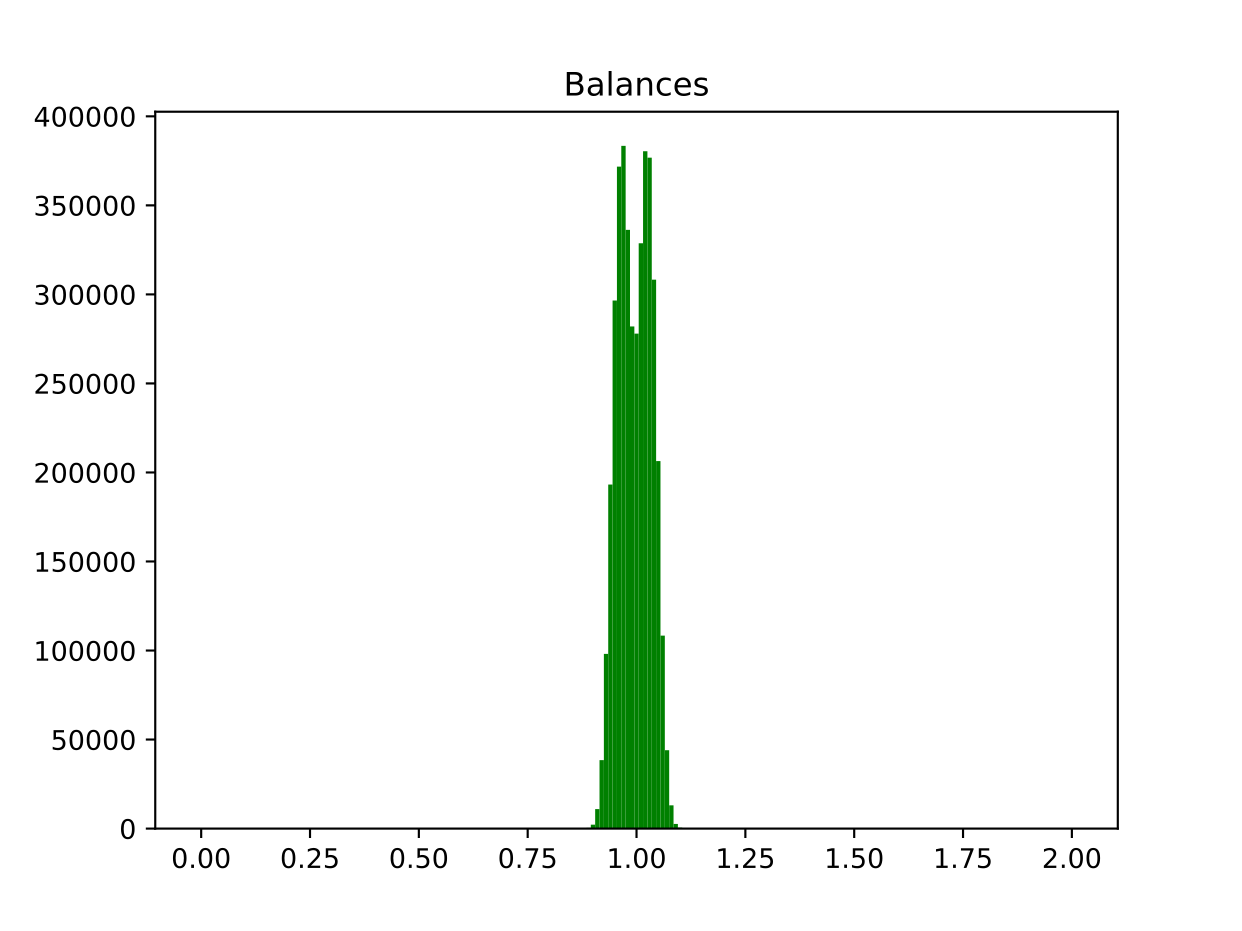}
    \end{tabular}
    \caption{The population balance histogram after running with balance constraints for steps starting from a vertical partition of a $40\!\times\!40$ grid graph. Since we start with a perfectly balanced partition, we do not include the first $1,000,000$ partitions in the histogram, and these figures were obtained after a month of computation. Left is subcritical ($\lambda = \nicefrac{1}{2\mu}$), middle is critical ($\lambda = \nicefrac{1}{\mu}$), right is supercritical ($\lambda = 1$). The balance constraints for the subcritical and supercritical runs are 50 percent, which explains the sharp cutoffs, and the balance constraint for the supercritical run is 100 percent.}
    \label{fig:popoverlap}
\end{figure}

The regions of apparent rapid and slow mixing, which depend on population balance and temperature, suggest that parallel tempering could be a useful way to sample from these stationary distributions. 
Parallel tempering between phase transitions probably would not work well, given that the corresponding distributions have negligible overlap in their support.
On the other hand, if one fixes $\lambda$ to be subcritical or critical and varies the population threshold $\alpha$, then there appears to be considerable overlap between the distributions; this is demonstrated in \Cref{fig:popoverlap}. Such overlap could be exploited by MCMC approaches using parallel tempering, as in \cite{Fifield_A_2018}. %

\subsection{Discussion}\label{sec:discussion}

\paragraph*{Mixing times.} %

Summarizing the data in \Cref{sec:metastableslopes} and \Cref{sec:SLEtests}, where we saw evidence about the interaction of the mixing time of flip walk with population constraints and fugacity regimes%
, we make the following conjectures. We recall that in the theoretical computer science literature, ``rapid'' mixing means polynomial mixing-time in the parameter, and ``torpid'' generally means super-polynomial or exponential time in the parameter. Here the parameter is $n$, the size of the grid graph. 

\begin{conjecture}\label{conj:mixingtimes}
Let $\mathcal{G} = (G_n)_{n \geq 0 }$ denote the family of $n \times n$ grid graphs. Then, there are functions $\epsilon_i : \mathbb{N} \to (0,1)$, $\epsilon_1 < \epsilon_2$ and $\lim_{n \to \infty} \epsilon_2(n) = 0$ so that the following holds. %
For $\lambda > 1 / \mu + \epsilon_2(n)$ the flip walk on $\mathcal{G}$ is torpidly mixing, regardless of population constraints. For $\lambda < 1/\mu - \epsilon_2(n)$, if there is any constant percentage population balance constraint then the flip walk on $\mathcal{G}$ is torpidly mixing, but with no population constraints it is rapidly mixing. For $|\lambda - 1/\mu| < \epsilon_1(n)$, the flip walk is rapidly mixing, regardless of of population constraints. 
\end{conjecture}

Although in \cite{najt2019complexity} we calculated worst case lower bounds on the mixing time of the flip walk, the graphs that provided the examples were unlike the graphs used in redistricting analysis. 
On the other hand, the grid graph seems to capture some of the difficulties involved in sampling connected partitions of the state dual graphs used in political districting. Thus resolving the above conjecture would bring us closer to understanding the reliability of flip-based Markov chains in redistricting. As previously mentioned, the flip walk appeared in statistical physics as the BFACF chain \cite{sokal1994monte}, but known lower bounds on mixing time \cite{sokal1989exponential} use techniques that we were not able to modify for this conjecture. In the remainder of this section we examine the applicability of annealing through different fugacities, the impact of fixing endpoints, present a schematic capturing some of the above conjecture and of the effects of annealing, and close by mentioning $k$-partitions briefly.

\paragraph*{Annealing.}

A common technique for escaping bottlenecks in MCMC is to use simulated annealing, where the inverse temperature parameter is varied to promote efficient movement throughout the state space. Although this technique has been used in court cases \cite{herschlag_evaluating_2017, herschlag_quantifying_2018}, the rigidity introduced by the contiguity and population balance terms for the redistricting problem mean that there are additional tradeoffs that must be evaluated in practice. %
\begin{figure}[!h]
    \centering
    \includegraphics[height=.5in]{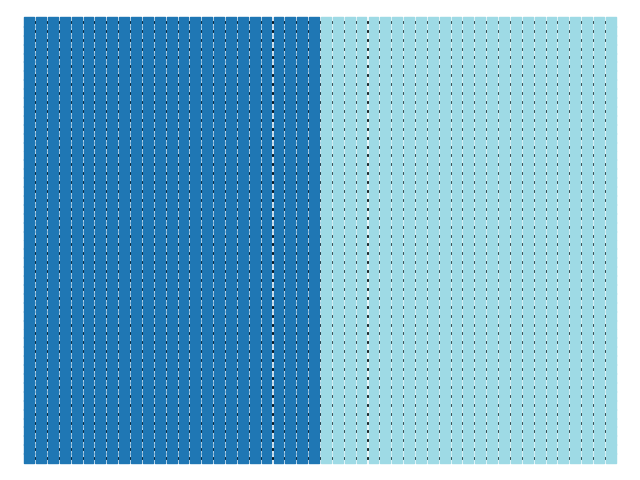}
    \includegraphics[height=.5in]{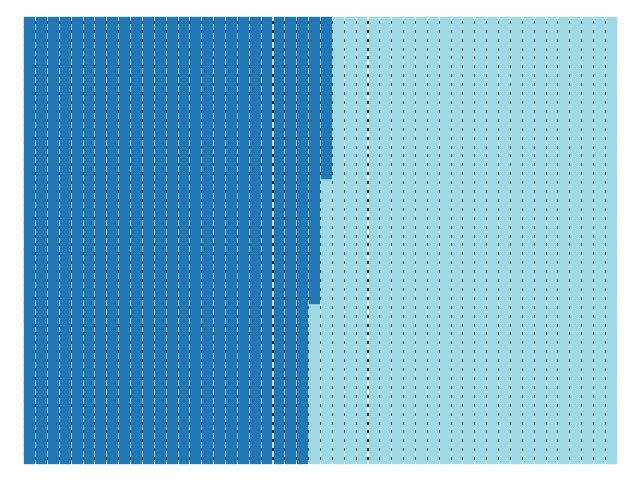}
    \includegraphics[height=.5in]{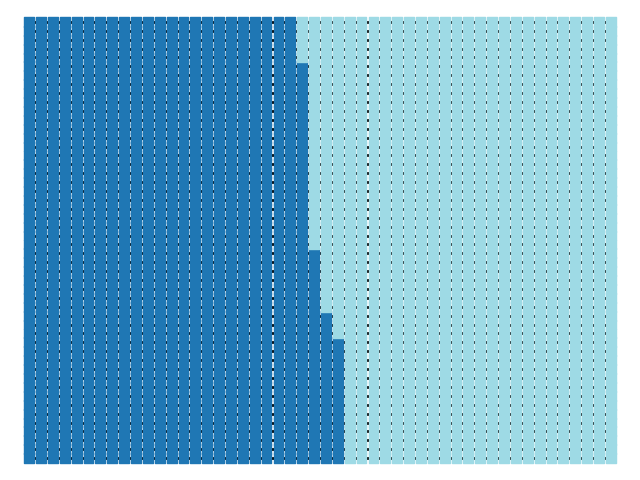}
    \includegraphics[height=.5in]{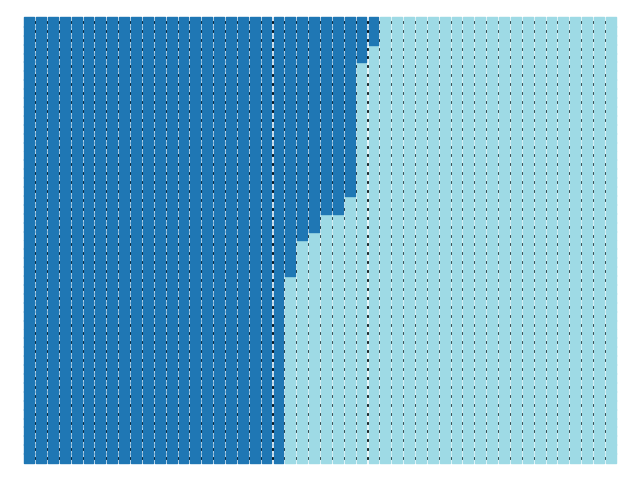}\\
        \includegraphics[height=.5in]{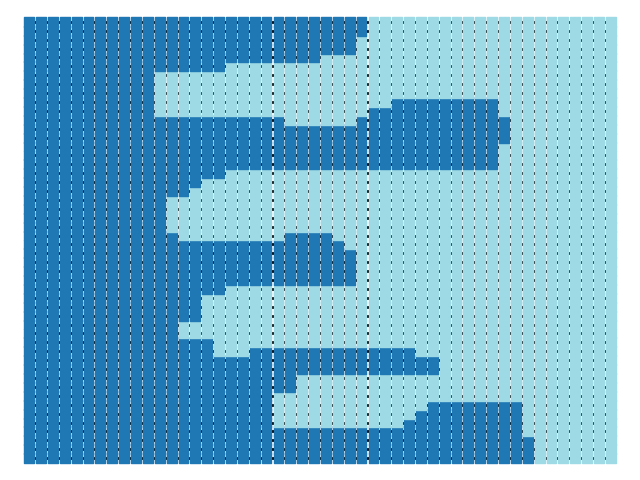}
    \includegraphics[height=.5in]{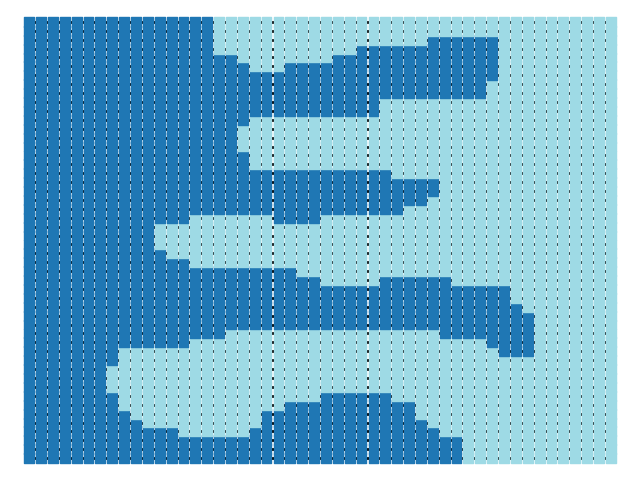}
    \includegraphics[height=.5in]{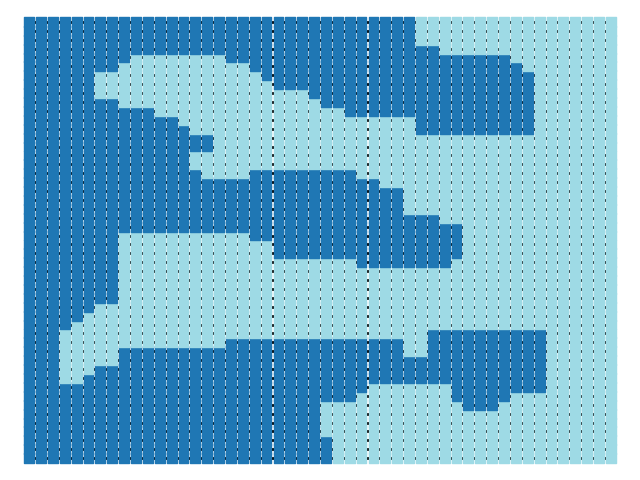}
    \includegraphics[height=.5in]{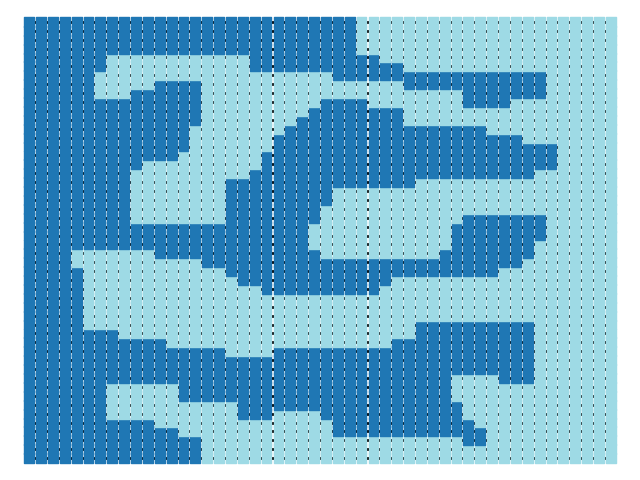}\\
    \includegraphics[height=.5in]{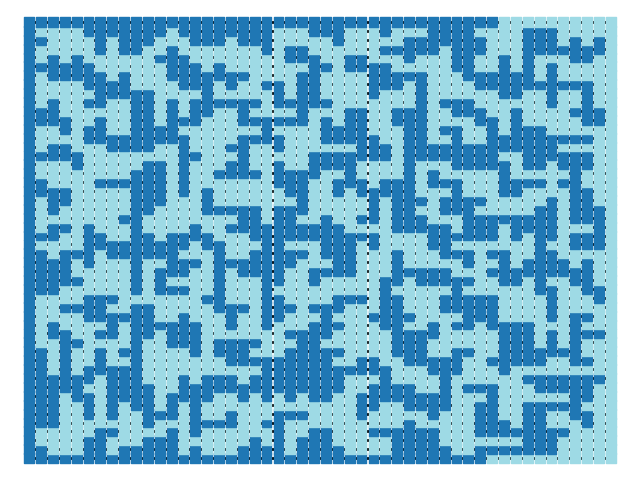}
    \includegraphics[height=.5in]{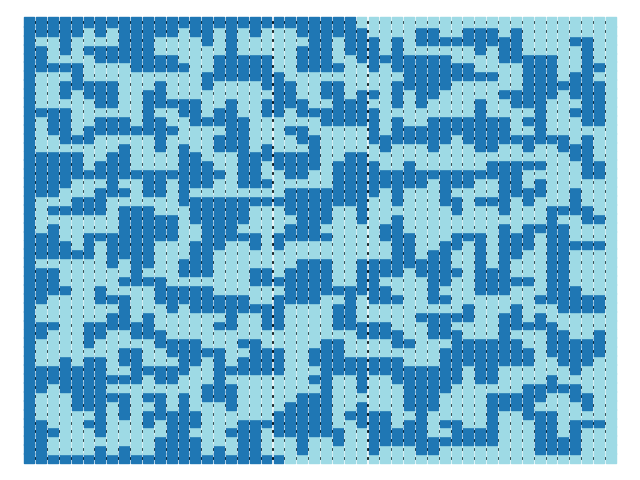}
    \includegraphics[height=.5in]{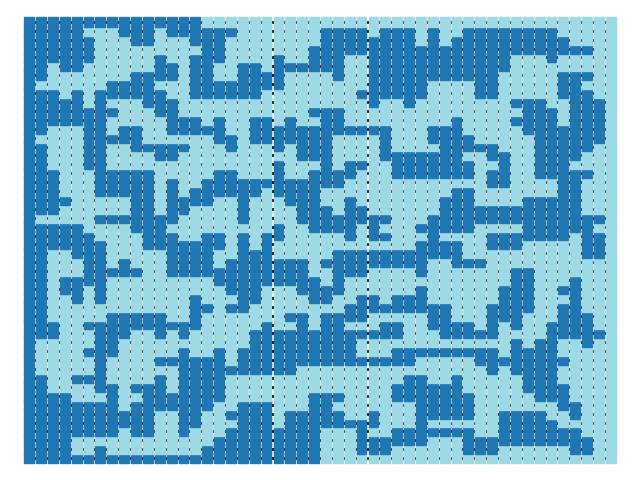}
    \includegraphics[height=.5in]{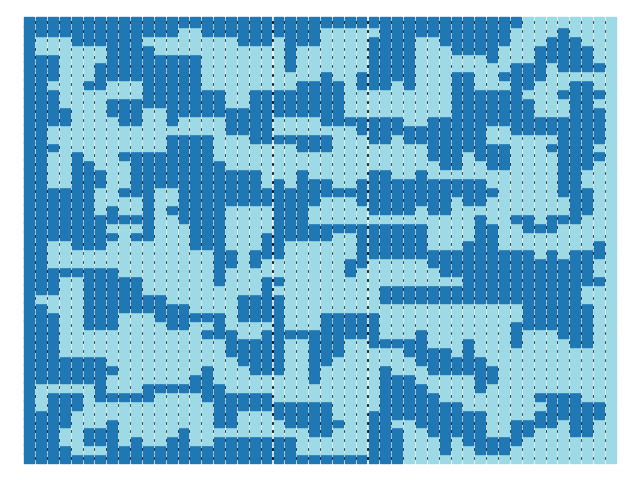}\\
    \includegraphics[height=.5in]{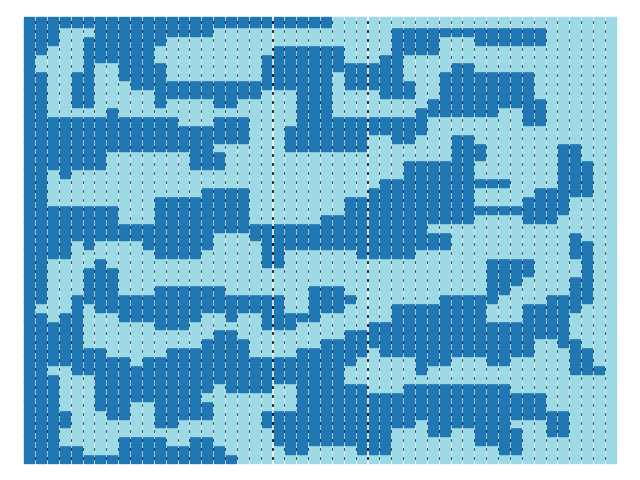}
    \includegraphics[height=.5in]{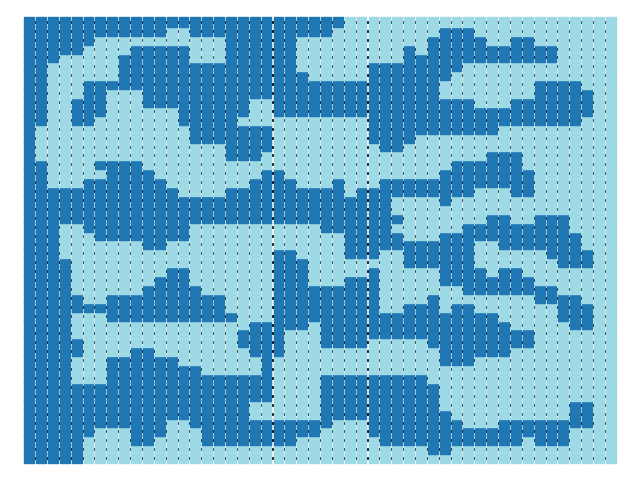}
    \includegraphics[height=.5in]{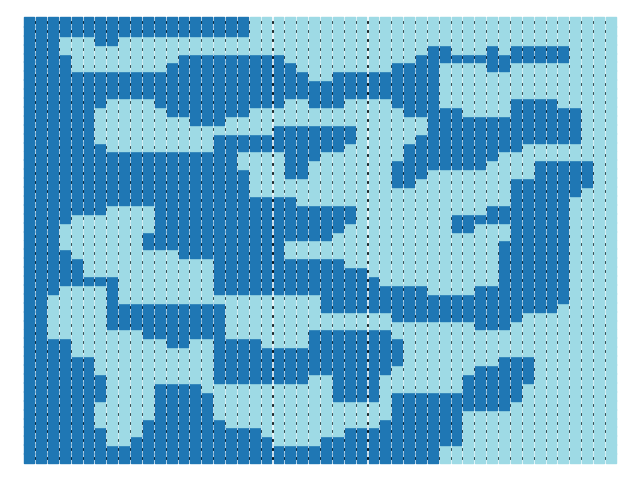}
    \includegraphics[height=.5in]{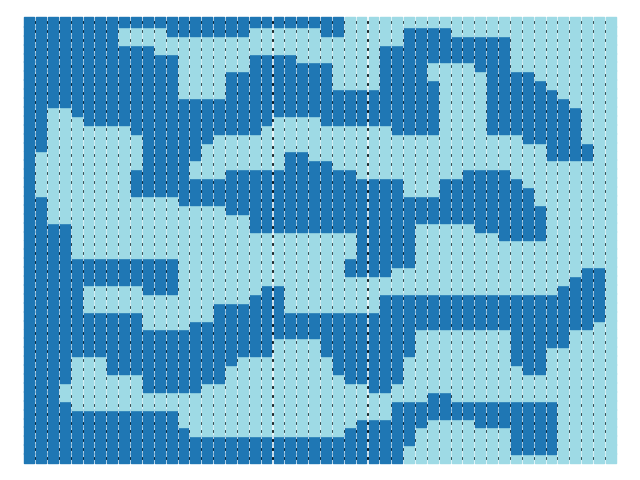}\\

    \caption{Samples from distributions with energy function $e^{-(\gamma \textrm{pop dev} + \delta \textrm{boundary length})}$
    with $\delta =1 $ as $\gamma$ varies from .11 to .81 (top two rows, read left to right) and $\gamma =1$  as $\delta$ varies from .11 to .81 (bottom two rows, read left to right). 
    The images in each experiment are snapshots of a single run of the Markov chain.}
    \label{fig:annealing}
\end{figure}

A key issue in many of the examples in \Cref{fig:annealing} is that while there is a significant amount of movement within the center of the grid, there is little change in the partition of the boundary component, which indicates that this simulated annealing strategy does not help us move out of the metastable region corresponding to the vertical partition. More explicitly, whether the induced partition of the boundary of the grid (a cycle graph) is vertical or horizontal decomposes the configuration space into two equal sized pieces, and \Cref{fig:annealing} shows that annealing does not essentially change the partition of that cycle graph---in fact, throughout the annealing process the boundary slope remains roughly vertical. A similar observation was made on real world data in \cite{deford2019recombination}.
Additionally, one source of computational inefficiency of these chains operating in the super-critical regime is that the space filling boundaries make it more likely that any individual proposal will be rejected since changing the assignment of a node is likely to disconnect the partition.

\paragraph*{Fixing endpoints. } %

As with the population balance constraint, there appears to be an appreciable difference between the mixing of connected partitions with fixed partition boundary endpoints, and where the partition boundary endpoints are allowed to move. Our empirical evidence does not exclude the possibility that the bottlenecks revealed by the boundary slope plots are the only bottlenecks in the super-critical and subcritical case; that would means that sampling from the stationary distribution could be possible by picking endpoints according to a partition function, and then running the flip walk on self-avoiding walks with fixed endpoints. Of course, such a strategy would become more difficult for $k$-partitions, where there is a combinatorial explosion in the different ways the endpoints of the boundary segments can be arranged.

\paragraph*{Space of graph configurations.} %

\begin{figure}[!h]

    \centering
    \graphicspath{{Images/schematic/}}
    \def\svgwidth{\columnwidth}
    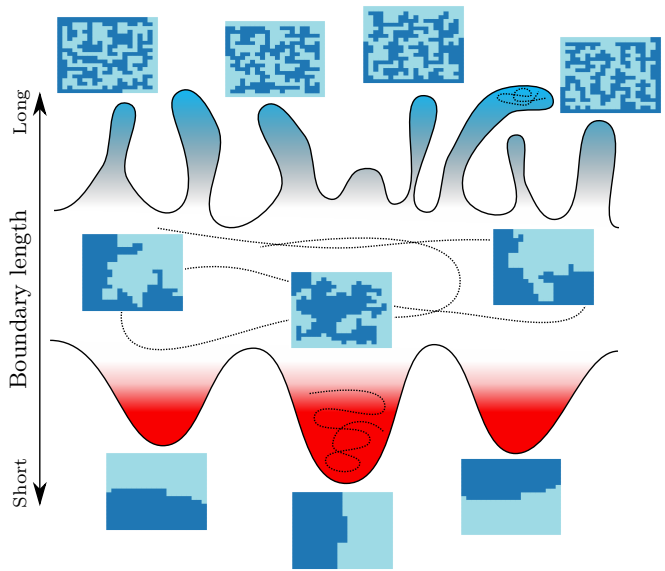

    \caption{A schematic showing the local geometry of connected graph partitions. The y-axis represents `compactness' (boundary length) as a measure of niceness of the partition.%
    }
    \label{fig:schematic}
\end{figure}

We give a schematic of the space of configurations in \Cref{fig:schematic}. As in our discussion of annealing, walks that attempt to travel vertically in our schematic by increasing or decreasing the fugacity do not appear to generate diverse collections of plans, since the boundary slopes remain essentially fixed. %
Traveling horizontally is also constrained outside of the critical region, as we saw earlier. %
In general, it appears that moving to other qualitatively distinct ``compact'' partitions requires traversing those with poorer compactness scores, such as those that are typical for the critical fugacity. On the other hand, allowing the compactness score to become too loose %
incurs a penalty in the form of the rigidity that comes from interlocking tendrils of supercritical fugacity states. Without population balance constraint, conditions are known for when these configurations spaces are connected \cite{akitaya2019reconfiguration}.  %

\paragraph*{$k$-partitions.}

An issue that arises when dealing with $k$-partitions instead of $2$-partitions is that compactness scores defined for individual districts have to be aggregated into compactness scores for an entire plan \cite{deford2019recombination,pegden1,herschlag_quantifying_2018}. %
There are various ways to perform this aggregation, and to the best of our knowledge phase transitions in choices of the aggregation have not been studied.

\section{The role of the state dual graph}\label{sec:role_of_discretization}

The experiments in \Cref{sec:phasetransitions} give an indication of the important role that the underlying graph plays in the way any given algorithm samples from the set of districting plans, thought of as partitions of the underlying geographic region. In this section, we highlight some subtle features of this dependence by showing that the behavior of a fixed sampling algorithm can be controlled by subdividing faces of the underlying state dual graph. We connect this to phase transitions in the self-avoiding walk model and discuss some implications for redistricting analysis.

\subsection{Concentration of boundary location}

\begin{figure}
    \centering
    \includegraphics[height = .8\linewidth, width =  .8\linewidth]{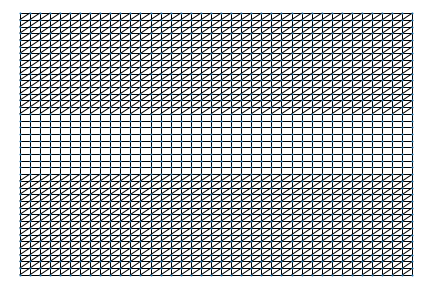}
    \caption{The graph $G'$.}
    \label{fig:metamandering_example}
\end{figure}

In this section we work with variations on the gridlandia graph, wherein we allow certain faces to be subdivided into triangles. 

\begin{figure}
    \centering
    \includegraphics[scale = .1, angle = 90]{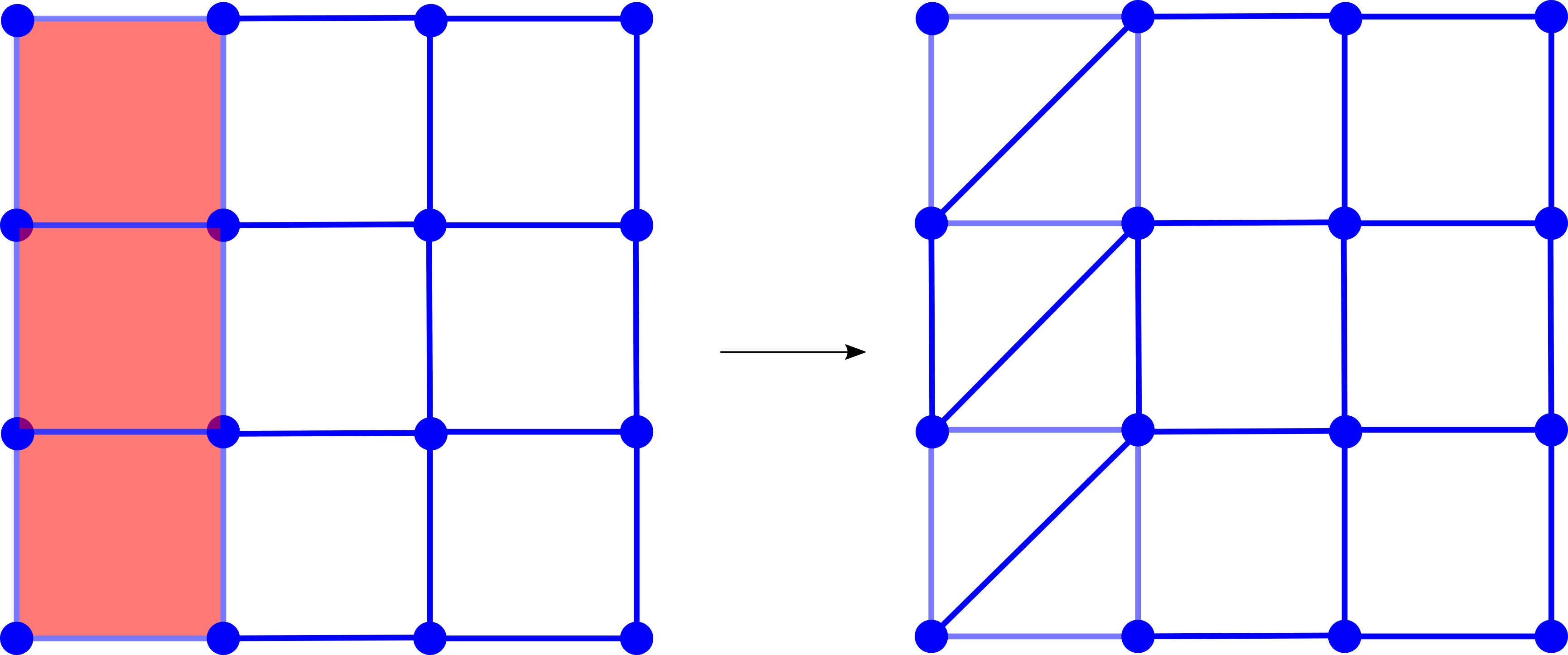}
    \caption{The shaded squares is the chosen set $F$, which are subdivided to form the graph on the right.}
    \label{fig:metamandering_subdivision_example}
\end{figure}

\begin{defn}[Subdivided Gridlandia]
Given gridlandia $G_n$, and a set $F \subseteq \Faces(G_n)$, where $\Faces$ denotes the set of bounded faces of $G_n$, define $G_n^F$ by adding edges diagonally from the lower left to upper right corner of each face in $F$. See \Cref{fig:metamandering_subdivision_example}.
\end{defn}

\begin{figure}
    \centering
\includegraphics[scale  =.1]{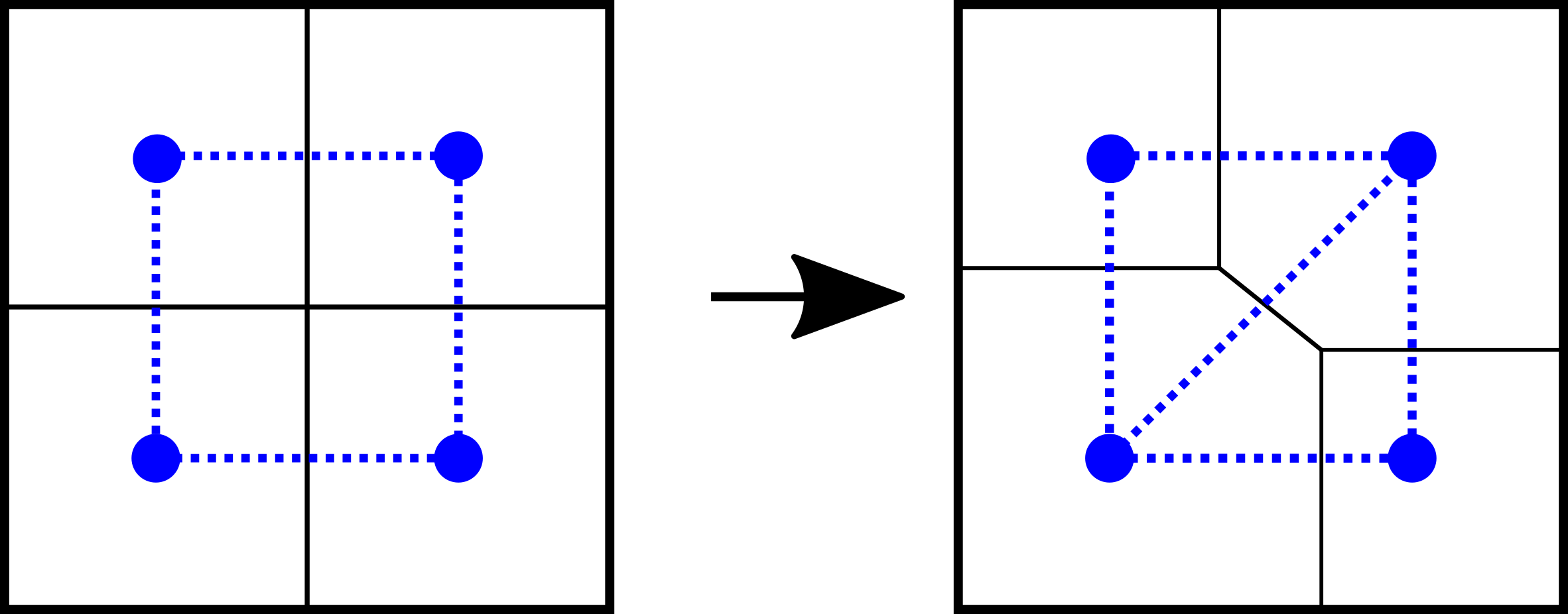}

    \caption{Minor modifications of the partition of a state into geographic units can change the lattice structure of the state dual graph.}
    \label{fig:how_to_metamander}
\end{figure}

\begin{remark}
$G_n$ is the dual graph of a map, that is, a partition of the unit square into regions wherein each region becomes a node, and two nodes are adjacent if the regions share a 1 dimensional boundary. In particular, $G_n$ is the dual graph of the map of the unit square obtained by subdividing it into regular squares, and $G_n^F$ is the dual graph obtained by modifying the partition of the square as in \Cref{fig:how_to_metamander}.
\end{remark}

Our aim here is to describe how carefully choosing $F$ can cause the boundary of ``random partition'' to concentrate around a horizontal line dividing gridlandia in half. We recall from \Cref{sec:phasetransitions} that the phase transition for connected partitions on the triangular lattice occurs at a lower fugacity than that of the square lattice. Thus, if parameters are set so the flip walk chain is subcritical on the triangular part, but critical on the square part, the chain will push partition boundaries off of the triangular portion. If the triangular portion surrounds the boundary of a particular partition $P$, it is therefore intuitive that that a stationary sample will be likely to thread its boundary along that of $P$.

\begin{figure}
    \centering
    \begin{tabular}{cc}
          \includegraphics[height = .5\linewidth, width = .5\linewidth]{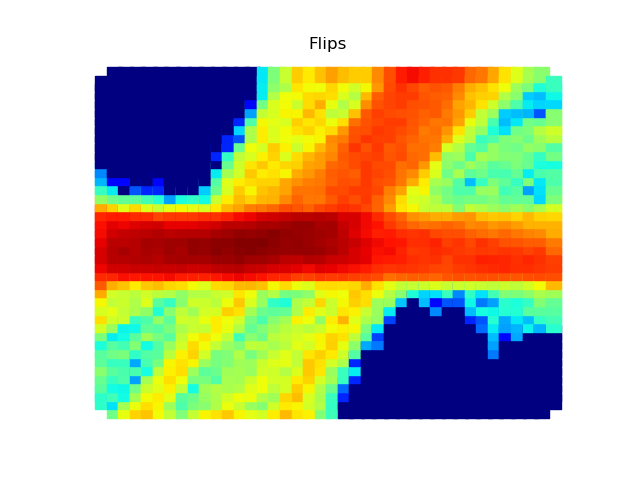}   &  \includegraphics[height = .5\linewidth, width = .5\linewidth]{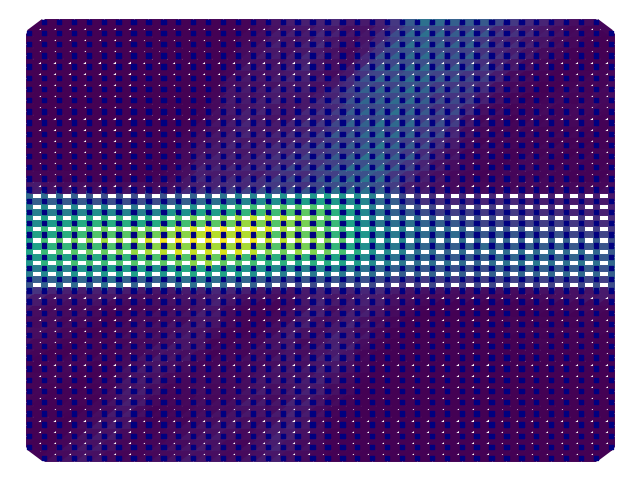}
    \end{tabular}

    \caption{Even if we start with the vertical partition, we rotate into the horizontal partition and then remain there. 1000000 steps. Logflips on the left and edge cuts heat map on the right.}
    \label{fig:rotation_metamandering_flip}
\end{figure}

\begin{figure}
    \centering
    \begin{tabular}{cc}
           \includegraphics[height = .5\linewidth, width = .5\linewidth]{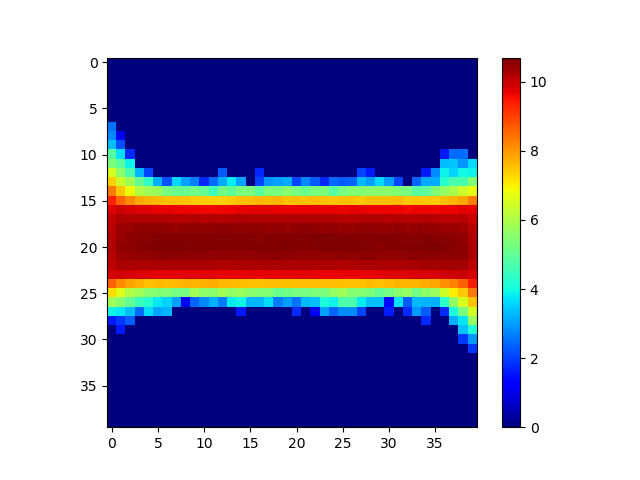}  & \includegraphics[height = .5\linewidth, width = .5\linewidth, angle = 90]{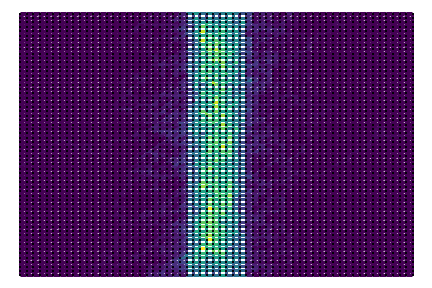}
    \end{tabular}

    \caption{ (Left) Log flips plot of running the flip walk with $\lambda = 1/\mu_{square}$ on $G'$ for 10,000,000 steps.  (Right) Edge cut heat map of splitting 1000 uniform spanning trees. In both cases a preference for the horizontal divide is apparent.}
    \label{fig:metamandering_images}
\end{figure}

This intuition is born out in the experiments we performed. Suppose we start with gridlandia $G_n$, but we triangulate the faces away from the horizontal partition boundary, as in \Cref{fig:metamandering_example}. We will call the resulting graph $G_n'$. We run the flip walk at fugacity $\lambda$ where $1/\mu_T < \lambda < 1/\mu_S$, and find a preference for partitions near to the horizontal partition. The result of this experiment can be seen in  \Cref{fig:metamandering_images}, where we can see that the boundary is indeed more likely to run vertically. This is a different feature from the subcritical metastable regions discussed in \Cref{sec:metastableslopes}, since a vertical partition of $G_n'$ will quickly rotate into being horizontal, see \Cref{fig:rotation_metamandering_flip}.

The typical behavior of partitions of $G_n'$ behaves similarly for the various popular sampling algorithms that partition by using components of random spanning forests \cite{deford2019recombination, carter2019merge} (see \Cref{sec:proposals}).  One such algorithm samples a uniform spanning tree and deletes a random edge such that the two components of the resulting forest are roughly balanced in size. When we sample connected partitions in this way and display the edge cut frequencies on the right side of \Cref{fig:metamandering_images}, there is a clear bias toward the horizontal partition. These biases can have significant effects on the number of seats won in a ``typical map,'' representing a challenge for ensemble based methods for curtailing extreme gerrymanders. Nearly identical effects are seen for producing a random spanning tree by drawing random edge weights and calculating the minimum spanning tree. See \cite{najt2019complexity} for further discussion, including an explicit example of the influence on the number of seats won. %

\begin{remark}
We explain a potential intuition for these examples. First, since we are partitioning the vertices of these graphs, we can represent the boundary by a simple cycle in the plane dual. A dual path passing through the triangulated part is in general twice as long as a corresponding boundary was before, since for every square face it passed through it now passes through two triangular faces. Even though exponentially many potential boundary paths are added after this triangulation, the compactness constraint is such that each is downweighted exponentially. The balancing act between having many paths and each path having negligible probability determines which orientation of boundary paths are more likely. While the phase transitions give a conceptual, although speculative, explanation of how this tug-of-war works out for the $\beta_{\lambda}$ distribution on the flip walk, it remains a challenge to explain the why $G'$ influences the uniform spanning tree based distribution in the way that it does.
\end{remark}
\begin{remark}
It is plausible that one could engineer a compactness score that depends on the metric length of the boundary, to avoid dependency on the discretization. %
However, defining such a compactness score raises several other important issues such as the fractal nature of coastlines, dependencies on map projections, and others \cite{duchin2018discrete, bar2019gerrymandering}. Investigating the relationship between state dual graph topology and compactness represents an important challenge for redistricting.
\end{remark}

\subsection{Implications for Redistricting}\label{sec:discretizations_implication}

We discuss the implications of the above experiments for redistricting analysis.

\paragraph*{Interpretation of the distribution.}

\begin{figure}
    \centering
        \includegraphics[height=.5in]{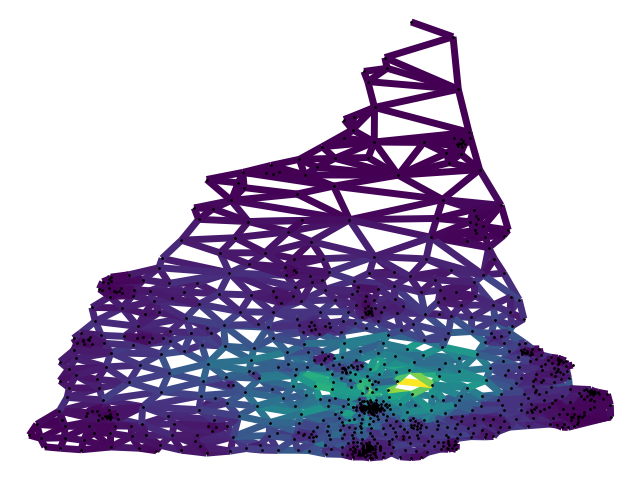}
    \includegraphics[height=.5in]{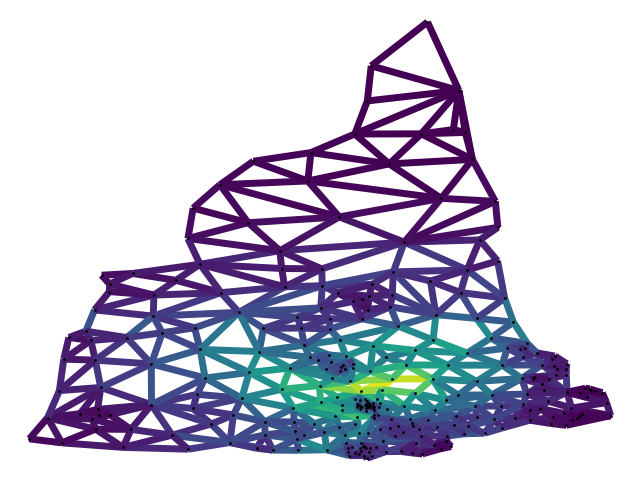}
    \includegraphics[height=.5in]{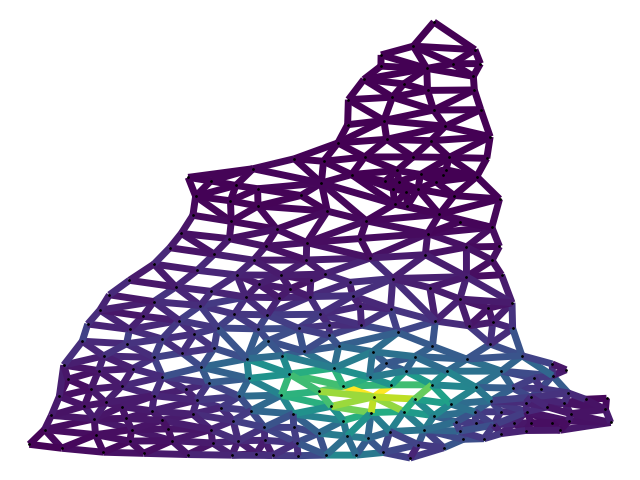}
    \includegraphics[height=.5in]{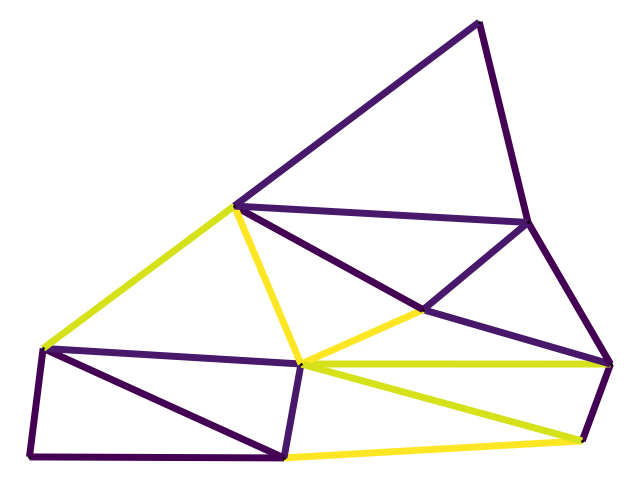}\\
    \includegraphics[height=.5in]{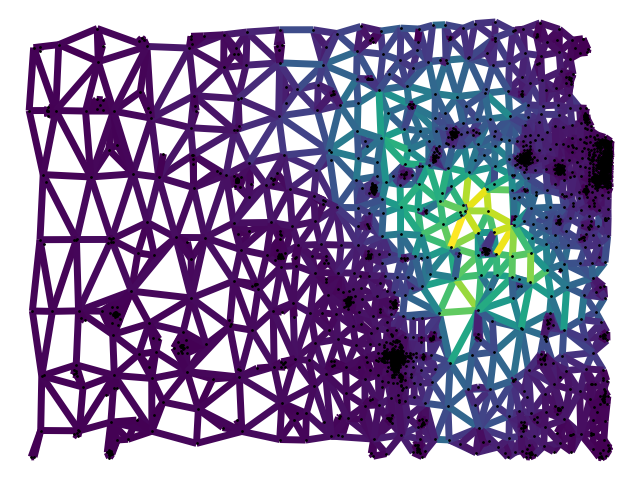}
    \includegraphics[height=.5in]{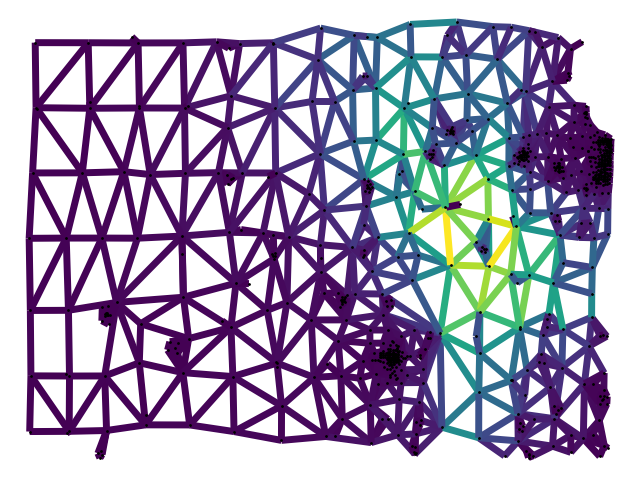}
    \includegraphics[height=.5in]{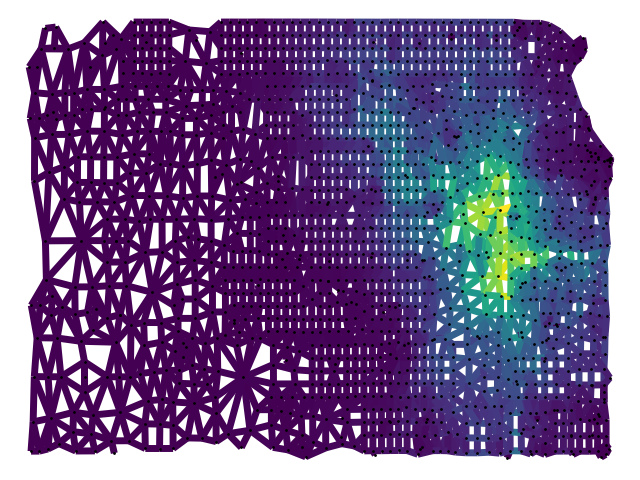}
    \includegraphics[height=.5in]{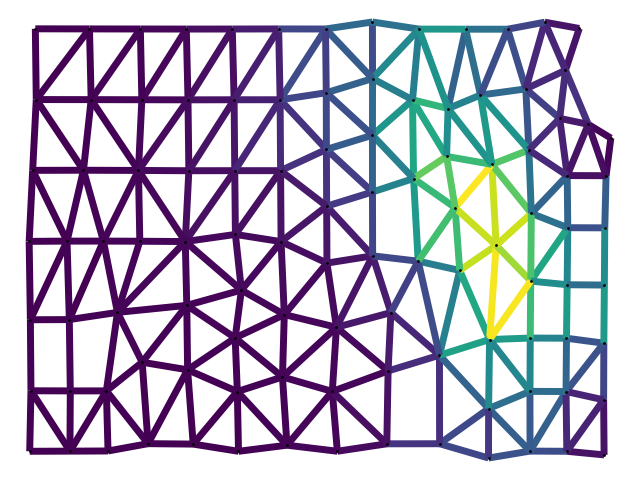}\\
    \caption{Heat map of boundary edge frequency for the states of New Hampshire and Kansas as computed over four different levels of resolution derived from census data. %
    For each state, we draw 10,000 districting plans, with a 2\% population deviation tolerance and compute how often each edge in the dual graph is cut in a partition.  }
    \label{fig:multiresolution}
\end{figure}

Recall that redistricting is modeled by producing a state dual graph out of small units. Common choices for these units include census geographies like blocks and tracts as well as voting precincts, whose boundaries are frequently determined locally. In \Cref{fig:multiresolution} we show the edge cut frequencies of spanning tree partitions for various choices of units. While the differences are not as extreme as the synthetic examples in \Cref{fig:metamandering_images}, they highlight the need for careful consideration of the underlying graph as a part of experimental design. There are many points in the analysis pipelines where choices about the state dual graph are made, such as handling state dual graph adjacencies over mountain ranges or lakes and prorating data between different geographic units. Thus, to rigorously compare plans to a typical plan it is important to understand to what extent the properties of typical plans depends on choices made earlier in the analysis, and to make such choices in a transparent and impartial way \cite{gelman2017beyond}.

\paragraph*{Tuning score functions.}

Investigating a local notion of the connective constant could lead to insights about a score function for which the flip walk mixes rapidly on a broad class of graphs. Following the definition of the connective constant in \cite{madrasself}, given a graph $G$ and $x \in V(G)$, we define $\mu_{G,n}(x) := \sqrt[n]{ S_n(G,x)}$, where $S_n(G,x)$ is the number of simple paths in $G$ starting at $x$ of length $n$. Then, for any self-avoiding walk $\gamma$ in $G$, defining $f_n(\gamma) := \prod_{x \in \gamma} \frac{1}{\mu_{G,n}(x)}$ and taking $n$ sufficiently large would give one possible analog to the $\beta_{1/\mu} (\gamma) = (1/\mu)^{|\gamma|}$ distribution, for which we gave evidence of rapid mixing in \Cref{sec:critical} and \Cref{sec:SLEtests}. A challenge for computing with this is that estimating $S_n(G,x)$ is not straightforward for large $n$ and is already challenging for highly structured graphs like $G = \mathbb{Z}^2$ \cite{clisby2013calculation}. %

\paragraph*{Metamandering.}

One could hypothetically exploit phase transitions and lattice-dependent properties to cause a sampling algorithm to have its sampling distribution concentrate around some preferred options, by an appropriate modification or subdivision of a state dual graph. For instance, if an analyst prefers vertical partitions of a square state, for instance because of the arrangement of voters, 
and had some freedom over the state dual graph, they could use the subdivided state dual graph to make a ``typical'' partition vertical. %
Whereas gerrymandering refers to the practice of drawing district boundaries to create desirable outcomes, we refer to practices that modify the input to a sampling algorithm to shift the seats histogram as \emph{metamandering}. Unlike gerrymandering, metamandering is more of academic interest than a practical concern, since scenarios allowing metamandering require far fetched and sophisticated conspiracies. Nevertheless, it raises a point of scrutiny in the analysis of districting plans by Markov chains. %

\section{Conclusion}

As methods from statistical physics continue to be incorporated into political redistricting analysis, domain knowledge of statistical physics becomes increasingly relevant to developing best practices. %
To support collaboration, we outline some challenging open questions at the interplay of redistricting and statistical physics.

Statistical physics motivates many questions in redistricting analysis, including the following:
\begin{enumerate}
    \item Are there phase transitions in the Gibbs energy scores mentioned in \Cref{sec:gibbs}? %
    \item Besides the $\beta_{\lambda}$ distributions, are there other analytically approachable distributions on the space of connected partitions of the grid graph?
    \item Are certain max entropy distributions over partitions, subject to certain partition parameters, efficiently sampleable?
    \item We know from statistical physics that there are certain features of models that are independent of the lattice being used. Although it would be surprising, perhaps there are similarly universal properties of outlier analysis? %
    \item Can we construct distributions over districting plans with fewer phase transitions or dependencies on discretization?
    \item How does the interaction between phase transitions, connective constants, and population balance influence mixing time of algorithms used to sample districting plans?
\end{enumerate}

Redistricting also motivates statistical physics style questions, such as:
\begin{enumerate}
    \item What is the influence of population balance on phase transitions in subcritical connected partitions of the grid?
    \item Is it possible to efficiently sample from self-avoiding walks in the $n \times n$ grid graph according to the $\beta_{\lambda}$ distribution? 
    \item When the endpoints are held fixed, does the flip walk mix rapidly at critical fugacity?
    \item What is the correct way to generalize the connective constant to irregular lattice-like graphs?
    \item How much of the population balance distribution in the critical fugacity case of \cref{fig:popoverlap} can calculated asymptotically via Schramm-Loewner evolution? See \cite{333336} for some discussion.
    \item Is a typical connected partition of the grid approximately balanced? See \cite{335793} for some discussion.
    \item Is there a scaling limit of the distribution of connected $k$-partitions of the $n \times n$ grid graph, with score function $\beta_{\lambda}(P) = \lambda^{\cut(P)}$ and $\lambda = \nicefrac{1}{\mu}$, where $\mu$ is the connective constant of the grid? What about just the $2$-partition case, with or without population balance? If so, can discretizations of it be sampled directly, as is the case with Schramm-Loewner evolution \cite{244227}? 
\end{enumerate}

\section{Acknowledgements}

We want to thank 
Assaf Bar-Natan, 
Sarah Cannon,
Jowei Chen, 
Moon Duchin,
Jordan Ellenberg, 
Gregory Herschlag, 
Matt Karrmann, 
Ian Gordon Mark, 
Mathoverflow user Kostaya$\_$I,
Jonathan Mattingly,
Marshall Mueller, 
Zohar Nussinov, 
John O'Neill, 
Wes Pegden, 
Dana Randall,
Zach Schutzmann,
Hannah Silverman,
Annie Yun, 
Paul Zhang,
and
Seanna Zhang 
for helpful discussions and/or feedback during the preparation of this manuscript.

L. Najt, D.\ DeFord and J.\ Solomon acknowledge the support of the Prof.\ Amar G.\ Bose Research Grant. J.\ Solomon acknowledges the support of Army Research Office grants W911NF1710068 and W911NF2010168, of Air Force Office of Scientific Research award FA9550-19-1-031, and of National Science Foundation grant IIS-1838071. L. Najt acknowledges the support of NSF TRIPODS Award $\#1740707$.

\bibliography{GerryRefs}

\end{document}